\documentclass[aps,prd,superscriptaddress,nofootinbib,eqsecnum,twocolumn]{revtex4-1}

\pdfoutput=1

\usepackage{amsfonts}
\usepackage{amsmath}
\usepackage{amssymb}
\usepackage{graphicx,color}
\usepackage{float}
\usepackage{hyperref}
\usepackage{subfigure}

\begin{document}

\title{Phase transition patterns for coupled complex scalar fields at finite temperature and density}

\author{Manuella C. Silva} \email{manuellacorrea13@gmail.com}
\affiliation{Departamento de F\'{\i}sica Te\'orica, Universidade do Estado do
  Rio de Janeiro, 20550-013 Rio de Janeiro, RJ, Brazil }

\author{Rudnei O. Ramos} \email{rudnei@uerj.br}
\affiliation{Departamento de F\'{\i}sica Te\'orica, Universidade do Estado do
  Rio de Janeiro, 20550-013 Rio de Janeiro, RJ, Brazil }
\affiliation{Physics Department, McGill University, Montreal, Quebec, H3A 2T8, Canada}

\author{Ricardo L. S. Farias} \email{ricardo.farias@ufsm.br}
\affiliation{Departamento de F\'{\i}sica, Universidade Federal de
  Santa Maria, Santa Maria, RS 97105-900, Brazil}

\begin{abstract}

The phase transition patterns displayed by a  model of two coupled
complex scalar fields  are studied at finite temperature and chemical
potential. Possible phenomena like symmetry persistence and inverse
symmetry breaking at high temperatures are analyzed. The effect of
finite density is also considered and studied in combination with the
thermal effects. The nonperturbative optimized perturbation theory
method is considered and the results contrasted with perturbation
theory. Applications of the results obtained are considered in the
context of an effective model for condensation of kaons at high
densities, which is of importance in the understanding of the
color-flavor locked phase of quantum chromodynamics.

\end{abstract}

\maketitle

\section{Introduction}
\label{intro}

In statistical physics, we can describe a dynamical system in the
ensemble as characterized by a Hamiltonian,  with charges (or quantum
numbers) and corresponding chemical potentials.  A chemical potential,
which enters as a Lagrange multiplier in the (quantum) grand-canonical
partition function, is assigned to  each conserved charge of the
system. This is the starting point of any setup aiming to study how a
conserved charge (or conserved charges) eventually affects the phase
structure of the system, e.g., in the Bose-Einstein condensation
problem~\cite{huang}.

The study of how a finite charge can affect the phase transition for a
scalar field in the context of quantum field theory dates back for
example from the pioneering works of Kapusta~\cite{Kapusta:1981aa},
Haber and Weldon~\cite{Haber:1981fg,Haber:1981ts} among others. In
Refs.~\cite{Bernstein:1990kf,Benson:1991nj} it was observed that
finite charges can modify strongly the phase transition structure of a
complex scalar field.  In particular, it was shown in details in
Ref.\cite{Benson:1991nj} that a sufficiently large fixed charge in the
context of a constant ratio for the number over entropy densities,
like for instance as expected to appear in cosmological settings, a
broken symmetry could persist at sufficiently high
temperatures. Likewise, under the same conditions, an originally
symmetric phase in the vacuum could get broken at high
temperatures. This is what characterizes a symmetry inversion phase
transition at high temperatures. These type of phenomena are
reminiscent of a symmetry nonrestoration (SNR) or inverse symmetry
breaking (ISB) type of transitions first studied by Weinberg in
Ref.~\cite{Weinberg:1974hy}.  The difference here, it is that these
unusual transition patterns studied in Ref.~\cite{Benson:1991nj} would
originate from finite density effects already in the case of an one
field case, whereas in Ref.~\cite{Weinberg:1974hy} they originate only
at finite temperatures for the case of multiple coupled scalar fields,
with both inter and intracouplings and with suitable choices of those
coupling constants. 

In the present paper, we reanalyze the effects of finite density in
the phase structure of complex scalar fields, but accounting for both
situations that were studied in Ref.~\cite{Weinberg:1974hy} and in
Ref.~\cite{Benson:1991nj}.  Here, we are then interested in the case
where the interplay of both coupling constants, thermal effects and
finite charges can compete or complement each other in the way they
can affect in unusual ways the phase structure of the system.  While
the finite charges effects in the phase structure of a complex scalar
field were originally studied in the context of the perturbative high
temperature approximation in Ref.~\cite{Benson:1991nj}, here we also
want to reevaluate that in terms of the nonperturbative method of the
optimized perturbation theory
(OPT)~\cite{Stevenson:1981vj,Okopinska:1987hp,Klimenko:1992av,Kleinert:1998zz,Chiku:1998kd,Pinto:1999py,Pinto:1999pg,Farias:2008fs}
(see also, e.g., Ref.~\cite{Yukalov:2019nhu} for a recent
review). This is specially important since perturbation theory studies
of phase transitions at high temperatures may be unreliable as it is
well known, which motivates the use of  different nonperturbative
methods (see, e.g.,
Refs.~\cite{Curtin:2016urg,Croon:2020cgk,Senaha:2020mop,Schicho:2021gca,Gould:2021oba}
for reviews and recent discussions).

At the final part of this work and as an applications of our results,
we will consider the condensation of kaons in the color-flavor locked
(CFL) phase of quantum chromodynamics (QCD).  The CFL phase is a
color-superconducting phase where diquark condensates break the chiral
symmetry~\cite{Alford:1998mk} (for a review, see, e.g.,
Ref.~\cite{Alford:2007xm}). The symmetry breaking pattern of this
phase transition can be associated with light pseudo-Nambu Goldstone
bosons, where the lightest of them are the charged and neutral
kaons. The study of the condensation of these light kaons has been
shown to be possible to be described in terms of an $O(2) \times
O(2)$-symmetric effective scalar field
theory~\cite{Alford:2007qa,Andersen:2008tn,Tran:2008mvg} that turns
out to be analogous to the coupled two complex scalal field we study
in the present paper.  In the previous
works~\cite{Alford:2007qa,Andersen:2008tn,Tran:2008mvg} the CFL phase
related to kaons condensation were studied in the
Cornwall-Jackiw-Tomboulis (CJT) nonperturbative
method~\cite{Cornwall:1974vz,Amelino-Camelia:1992qfe}.  The use of the
OPT method here allows us not only to compare with those earlier
results obtained with the CJT method, but also allows us to discuss
the Goldstone theorem, which was shown to be problematic in those
studies. However,  in the OPT case, the Goldstone theorem is exactly
satisfied as we will show.

The remainder of this paper is organized as follows. In
Sec.~\ref{section2}, we introduce the model studied in this paper,
along also with its implementation in the context of the OPT
method. The relevant thermodynamic quantities for our study are also
derived. In Sec.~\ref{results}, we give the many numerical results
exploring both ISB and SNR realizations in the model and the many
possibilities of phase transition patterns that can emerge at finite
chemical potential and charge densities in a thermal environment. In
Sec.~\ref{kaons}, we apply our results for the case of an effective
model derived from a chiral Lagrangian density describing the
condensation of kaons  in a CFL phase for QCD at high densities. Our
conclusions are presented in Sec.~\ref{conclusions}.  Two appendices
are also included, where some of the technical details are presented.

\section{OPT implementation for the two complex scalar field model}
\label{section2}

We consider a model with two complex scalar fields, $\phi$ and $\psi$,
with quartic self-interactions and a biquadratic intercoupling
between them.  The Lagrangian density is given by
\begin{eqnarray}
\mathcal{L} &=&  ( \partial_{\mu }\phi) (\partial^{\mu }\phi^{*})-
m_{\phi }^{2}(\phi \phi^{*}) -\frac{\lambda_{\phi }}{6} (\phi
\phi^{*})^{2}  \notag \\ &-& ( \partial_{\mu }\psi) (\partial^{\mu
}\psi^{*})  -m_{\psi }^{2}(\psi \psi^{*})  -\frac{\lambda_{\psi }}{6}
(\psi \psi^{*})^{2} \nonumber \\ &-&\lambda
(\phi\phi^{*})(\psi\psi^{*}).
\label{lagrangian}
\end{eqnarray}

It is convenient to write the complex scalar fields $\phi$ and $\psi$
in terms of their real and imaginary components as, $\phi = (\phi_1+i
\phi_2)/\sqrt{2}$ and $\psi = (\psi_1+i \psi_2)/\sqrt{2}$. The
Lagrangian density (\ref{lagrangian}) then becomes
\begin{eqnarray}
\mathcal{L} &=& \frac{1}{2}\left(\partial_{\mu }\phi_1 \right)^2 +
\frac{1}{2}\left(\partial_{\mu }\phi_2 \right)^2 \nonumber \\ &+&
\frac{1}{2}\left(\partial_{\mu }\psi_1 \right)^2
+\frac{1}{2}\left(\partial_{\mu }\psi_2 \right)^2  \nonumber \\ &-& V,
\label{lagr2}
\end{eqnarray}
where the tree-level potential $V$ is
\begin{eqnarray}
V &=&  \frac{ m_{\phi }^2}{2} \left(\phi_1^2+\phi_2^2\right) +
\frac{m_{\psi}^2}{2} \left(\psi_1^2+\psi_2^2\right)  \nonumber \\ &+&
\frac{\lambda_{\phi }}{4!} \left(\phi_1^2+\phi_2^2\right)^2
+\frac{\lambda_{\psi}}{4!} \left(\psi_1^2+\psi_2^2\right)^2  \nonumber
\\ &+& \frac{\lambda}{4}\left(\phi_1^2+\phi_2^2\right)
\left(\psi_1^2+\psi_2^2\right).
\label{potential}
\end{eqnarray}
The stability of the potential requires $\lambda_\phi>0$,
$\lambda_\psi >0$ and, when $\lambda<0$, that $\lambda_\phi
\lambda_\psi > 9 \lambda^2$. The intercoupling $\lambda$, hence, can
be either positive or negative. We will be particularly interested in
the case where $\lambda$ can assume negative values.

\subsection{The effective potential at finite temperature and chemical potential}

The thermodynamical potential density is defined as
\begin{equation}
V_{\rm eff} (\mu, T) = -\frac{T}{{\cal V}} \ln Z(\beta,\mu),
\label{Vthermo}
\end{equation}
where ${\cal V}$ is the space volume and $Z(\beta,\mu)$ is the grand
partition function,
\begin{equation}
Z(\beta,\mu) = {\rm tr} \left[e^{-\beta ({\hat H} - \mu_a {\hat
      Q}_a)}\right],
\label{Zbetamu}
\end{equation}
where $\beta=1/T$, ${\hat H}$ is the Hamiltonian operator, ${\hat
  Q}_a$ denotes the conserved charge operators and $\mu_a$ are the
corresponding chemical potentials. The grand partition function can be
written as a  functional integral over the fields as usual in quantum
field theory~\cite{Kapusta:2006pm}.  In the integral functional field
form, the grand partition function for the model
Eq.~(\ref{lagrangian}) then becomes
\begin{eqnarray}
Z(\beta,\mu) = \int_{\rm periodic} D\phi_1 D\phi_2 D\psi_1 D\psi_2\,
e^{-S_{\rm Eucl}},
\label{Zfunct}
\end{eqnarray}
where the functional integrals over the fields are performed under the
periodic boundary  conditions in imaginary Euclidean time,
$\phi_i({\bf x},0)= \phi_i({\bf x},\beta)$ and  $\psi_i({\bf x},0)=
\psi_i({\bf x},\beta)$, and the Euclidean action $S_{\rm Eucl}$ is
given by
\begin{eqnarray}
S_{\rm Eucl} &=& \int_0^\beta d \tau \int d^3 x \left[ \frac{1}{2}
  \left( \frac{\partial \phi_1}{\partial \tau} - i \mu_\phi
  \phi_2\right)^2 \right.  \nonumber \\ &+& \left. \frac{1}{2} \left(
  \frac{\partial \phi_2}{\partial \tau} + i \mu_\phi \phi_1\right)^2
  \right.  \nonumber \\ &+& \left. \frac{1}{2}(\nabla\phi_1)^2 +
  \frac{1}{2}(\nabla\phi_2)^2 + \frac{m_\phi^2}{2} (\phi_1^2 +
  \phi_2^2) \right.  \nonumber \\ &+& \left. \frac{\lambda_\phi}{24}
  (\phi_1^2 + \phi_2^2)^2 \right.  \nonumber \\ &+& \left. \frac{1}{2}
  \left( \frac{\partial \psi_1}{\partial \tau} - i \mu_\psi
  \psi_2\right)^2 + \frac{1}{2} \left( \frac{\partial \psi_2}{\partial
    \tau} + i \mu_\psi \psi_1\right)^2 \right.  \nonumber \\ &+&
  \left. \frac{1}{2}(\nabla\psi_1)^2 +  \frac{1}{2}(\nabla\psi_2)^2 +
  \frac{m_\psi^2}{2} (\psi_1^2 + \psi_2^2) \right.  \nonumber \\ &+&
  \left. \frac{\lambda_\psi}{24} (\psi_1^2 + \psi_2^2)^2 \right.
  \nonumber \\ &+&\left. \frac{\lambda}{4}
  \left(\phi_1^2+\phi_2^2\right)\left(\psi_1^2+\psi_2^2\right)
  \right].
\label{SEucl}
\end{eqnarray}
The expression (\ref{Zfunct}) with Eq.~(\ref{SEucl}) generalizes for
the present case of a system with two complex coupled scalar fields
the one obtained in the case of one only complex scalar field case, as
given, e.g., in Ref.~\cite{Kapusta:2006pm} (see also
Ref.~\cite{Benson:1991nj}).

\subsection{The OPT implementation}

The general implementation of the OPT in the Lagrangian density
happens through an interpolation defined as (see, e.g.,
Refs.~\cite{Rosa:2016czs,Farias:2021ult} and references there in)
\begin{equation}
  \mathcal{L} \rightarrow\mathcal{L}^{\delta} =
  (1-\delta)\mathcal{L}_{0}(\eta) + \delta \mathcal{L} , 
\label{inter}
\end{equation}
where $\mathcal{L}_{0} $ is the Lagrangian density of the free theory,
which is modified by an arbitrary mass parameter $\eta$. 
 
The standard interpolation procedure given by Eq.~(\ref{inter}) gives
for our model the following  Lagrangian density
\begin{eqnarray}
\mathcal{L}^{\delta} &=&\frac{1}{2}\left[ \left( \partial_{\mu
  }\phi_{1}\right)^{2} +\left( \partial_{\mu
  }\phi_{2}\right)^{2}+\left( \partial_{\mu}\psi_{1}\right)^{2}
  +\left( \partial_{\mu }\psi_{2}\right)^{2}\right] \nonumber \\ &-&
\frac{m_{\phi }^{2}}{2}\left( \phi_{1}^{2}+\phi_{2}^{2}\right)
-\frac{m_{\psi }^{2}}{2}\left( \psi_{1}^{2}+\psi_{2}^{2}\right)
\nonumber \\ &-&\frac{\delta}{4!}\lambda_{\phi }\left(
\phi_{1}^{2}+\phi_{2}^{2}\right)^{2} -\frac{\delta}{4!}\lambda_{\psi
}\left( \psi_{1}^{2}+\psi_{2}^{2}\right)^{2} \nonumber
\\ &-&\frac{\delta}{4}\lambda \left( \phi_{1}^{2}+\phi_{2}^{2}\right)
\left(\psi_{1}^{2}+\psi_{2}^{2}\right)   \nonumber \\ &-&\left(
1-\delta \right) \frac{\eta_{\phi
  }^{2}}{2}\left(\phi_{1}^{2}+\phi_2^2\right) -\left( 1-\delta\right)
\frac{\eta_{\psi }^{2}}{2}\left(\psi_{1}^{2}+\psi_2^2\right) \nonumber
\\ &=&\frac{1}{2}\left[ \left( \partial_{\mu
  }\phi_{1}\right)^{2}+\left( \partial_{\mu
  }\phi_{2}\right)^{2}+\left( \partial_{\mu }\psi_{1}\right)^{2}
  +\left( \partial_{\mu }\psi_{2}\right)^{2}\right]  \nonumber
\\ &-&\frac{\Omega_{\phi }^{2}}{2}\left(
\phi_{1}^{2}+\phi_{2}^{2}\right)  -\frac{\Omega_{\psi }^{2}}{2} \left(
\psi_{1}^{2}+\psi_{2}^{2}\right)   \nonumber \\ &-&\frac{1}{4!}\delta
\lambda_{\phi }\left( \phi_{1}^{2}
+\phi_{2}^{2}\right)^{2}-\frac{1}{4!}\delta \lambda_{\psi }\left(
\psi_{1}^{2}+\psi_{2}^{2}\right)^{2} \nonumber \\ &-&\frac{1}{4}\delta
\lambda \left( \phi_{1}^{2}+\phi_{2}^{2}\right) \left(
\psi_{1}^{2}+\psi_{2}^{2}\right) \nonumber \\ &+&\frac{1}{2}\delta
\eta_{\phi }^{2}\left( \phi_{1}^{2}+\phi_{2}^{2}\right)
+\frac{1}{2}\delta \eta_{\psi }^{2}\left(
\psi_{1}^{2}+\psi_{2}^{2}\right) ,
\label{OPTL}
\end{eqnarray}
with $\Omega_{\phi }^{2}=m_{\phi }^{2}+\eta_{\phi }^{2}$,
$\Omega_{\psi }^{2}=m_{\psi }^{2}+\eta_{\psi }^{2}$, where
$\eta_{\phi,\psi}$ are mass parameters determined through a
variational procedure (see below) and $\delta$ is a dimensionless
parameter used as a bookkeeping parameter only to keep track of the
order that the OPT is implemented and $\delta$  is set to one at the
end. Note that the OPT interpolation changes the usual {}Feynman rules
of the theory.  The quartic vertices are changed to
\begin{eqnarray}
&& -i \lambda_\phi \to -i \delta \lambda_\phi, \nonumber \\ && -i
  \lambda_\psi \to -i \delta \lambda_\psi, \nonumber\\ && -i \lambda
  \to -i \delta \lambda,
\label{vertices}
\end{eqnarray}
and the OPT also leads to the additional vertices that come from the
interpolation procedure and that are quadratic in the fields, which
comes from the last two terms appearing in Eq.~(\ref{OPTL}).  Their
{}Feynman rules are simply
\begin{eqnarray}
i \delta \eta_\phi^2 \; \; \; {\rm and}\;\; i \delta \eta_\psi^2,
\label{etavertices}
\end{eqnarray}
while the bare propagators for the fields have masses replaced in the
OPT procedure by $m_\phi^2 \to \Omega_{\phi }^{2}$ and $m_\psi^2 \to
\Omega_{\psi }^{2}$.

All calculations are carried out similarly as done in perturbation
theory and can be evaluated at any order in $\delta$. Hence, up to
this stage the results remain strictly perturbative and very similar
to the ones obtained via an ordinary perturbative calculation.  Since
all quantities evaluated at any finite order $\delta^k$ in the OPT
depends explicitly on $\eta_{\phi, \psi }$, these parameters need to
be fixed appropriately.  It is through the freedom in fixing
$\eta_{\phi, \psi }$ that nonperturbative results can be generated in
the OPT. Since $\eta_{\phi, \psi }$ do not belong to the original
theory, these parameters need to be fixed considering some appropriate
procedure.  {}For instance, one may fix them by requiring that a given
physical quantity, which is been calculated perturbatively to
order-$\delta^{k}$, to be evaluated at the value where it is less
sensitive to this parameter.  This criterion is known as the principle
of minimal sensitivity (PMS)~\cite{Stevenson:1981vj}.  In this work we
consider the effective thermodynamic potential (ETP), which is
evaluated to  order-$\delta^{k}$, $V_{\rm eff}^{\left( \delta^k\right)
}$, as the appropriate quantity to be optimized and as considered in
most of the OPT works in general\footnote{See, for example,
  Refs.~\cite{Farias:2008fs,Rosa:2016czs,Yukalov:2019nhu}, for
  examples of other different quantities that can be optimized,
  different optimization methods and a comparison between the
  results.}. In this case, the PMS criterion translates into the
variational relation
\begin{equation}
\frac{\partial V_{\rm eff}^{( \delta^k ) }}{\partial\eta_{\phi, \psi
}}\Bigr|_{\eta_{\phi, \psi }=\bar{ \eta}_{\phi, \psi },\delta
  =1}=0.  \label{pms}
\end{equation}
There can also be other optimization procedures that can be applied to
fix the OPT mass parameters,  but they have shown to be equivalent to
the PMS one, including the convergence properties (see,  see for
instance, Ref.~\cite{Rosa:2016czs} for a discussion on these issues).
The optimum value for $\bar{\eta}_{\phi}$ and $\bar{\eta}_{ \psi }$
derived from Eq.~(\ref{pms}) are now nontrivial functions of the
original parameters of the theory. In particular, $\bar{\eta}_{\phi}$
and $\bar{\eta}_{ \psi }$ become explicit functions of the couplings
and, as a consequence of this, nonperturbative results are generated.

\subsection{The ETP in the OPT approximation}
\label{freeEnergyTB}

\begin{figure}[!htb]
\centerline{ \includegraphics[width=7.5cm]{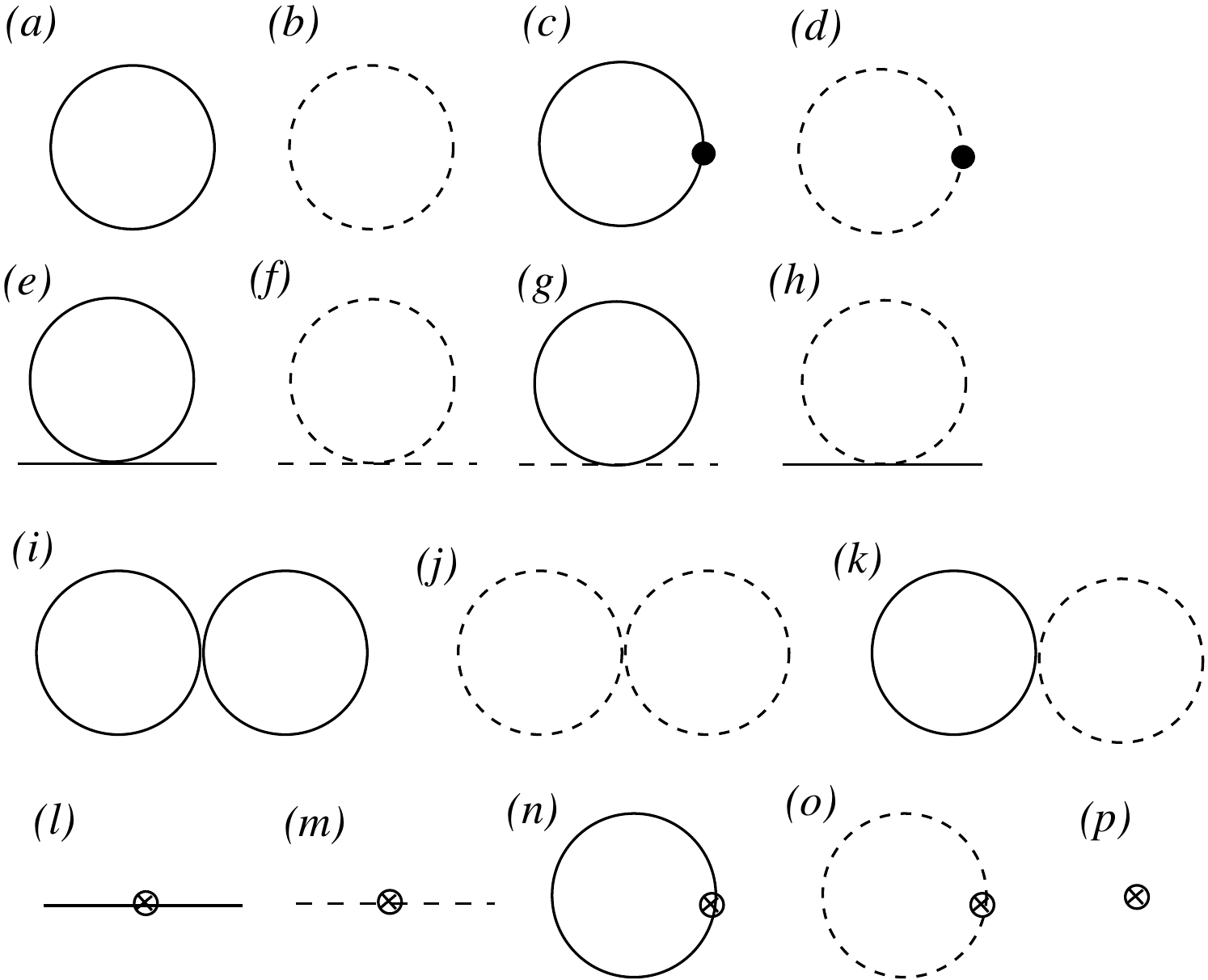}}
\caption{All Feynman diagrams contributing to the ETP up to first
  order in the OPT.  External lines refer to insertions of the scalar
  background fields $\phi_0$ and $\psi_0$. Solid and dashed lines
  stand for $\phi $ and $ \psi $ propagators, respectively. A black
  dot is an insertion of $ \delta \eta^{2}_{\phi,\psi}$ [see
    Eq.~(\ref{etavertices})].  The terms (l) and (m) indicate the mass
  renormalization counterterms, with (n) and (o) showing the
  corresponding diagrams constructed from these mass counterterms at
  order $\delta$.  {}Finally, the last term (p) denotes a simple
  vacuum renormalization counterterm at order $\delta$.}
\label{fig1}
\end{figure}

By shifting the fields around their respective background expectations
values in Eq.~(\ref{OPTL}),  which can be taken along $\phi_1$ and
$\psi_1$ without loss of generality, then $\phi_1 \to \phi_1^\prime =
\phi_1 + \phi_0$ and $\psi_1 \to \psi_1^\prime = \psi_1 + \psi_0$,
with $\langle \phi_1 \rangle = \langle \phi_2 \rangle = \langle \psi_1
\rangle = \langle \psi_2 \rangle = 0$ and  $\langle \phi_1^\prime
\rangle = \phi_0$ and $\langle \psi_1^\prime \rangle = \psi_0$, we
can, thus, derive the effective potential at first order in the OPT,
i.e., at first order in $\delta$.  All terms contributing to the ETP
to first order in the OPT are given in {}Fig.~\ref{fig1}. They are all
explicitly derived in the Appendix~\ref{appA} and  the renormalized
ETP at first order in the OPT is then given by
\begin{eqnarray}
V_{\rm eff,R}^{(\delta)} &=&
\frac{m_\phi^2-\mu_\phi^2}{2}\phi_0^{2}+\frac{m_\psi^2-\mu_\psi^2}{2}\psi_0^{2}
\nonumber \\ &+&\frac{\lambda_{\phi }}{4!} \phi_0^{4}+\frac{
  \lambda_{\psi }}{4!}\psi_0^{4}+\frac{ \lambda}{4}
\phi_0^{2}\psi_{0}^{2} \nonumber \\ &+& Y(\Omega_\phi,T,\mu_\phi)+
Y(\Omega_\psi,T,\mu_\psi) \nonumber\\ &+& \left(\frac{\lambda_{\phi
}}{3} \phi_0^{2} +\frac{\lambda }{2} \psi_0^{2}  -\eta_{\phi
}^{2}\right)      X(\Omega_\phi,T,\mu_\phi) \nonumber \\ &+&
\left(\frac{\lambda_{\psi }}{3} \psi_0^{2} + \frac{\lambda }{2}
\phi_0^{2}-\eta_{\psi }^{2} \right)     X(\Omega_\psi,T,\mu_\psi)
\nonumber \\ &+& \frac{\lambda_{\phi}}{3} X^2(\Omega_\phi,T,\mu_\phi)
+\frac{\lambda_{\psi}}{3} X^2(\Omega_\psi,T,\mu_\psi) \nonumber \\ &+&
\lambda X(\Omega_\phi,T,\mu_\phi) X(\Omega_\psi,T,\mu_\psi),
\label{VeffR}
\end{eqnarray}
where the functions $Y(\Omega_i,T,\mu_i)$ and $X(\Omega_i,T,\mu_i)$
haven been defined in the Appendix~\ref{appA} and given by
Eqs.(\ref{YTmu}) and (\ref{XTmu}), respectively.

\subsection{Optimization procedure}

Applying the PMS procedure Eq.~(\ref{pms}) to the renormalized ETP
Eq.~(\ref{VeffR}), we obtain that $\bar{\eta}_\phi$ and
$\bar{\eta}_\psi$ are obtained from the coupled equations,
\begin{eqnarray}
\bar{\eta}_\phi^{2}&=&\frac{\lambda_\phi
}{3}\tilde{\phi}^2+\frac{\lambda}{2} \tilde{\psi}^2 \nonumber
\\ &+&\frac{2\lambda_\phi }{3}
X(\Omega_\phi,T,\mu_\phi)\Bigr|_{\eta_\phi=\bar{\eta}_\phi}+ \lambda
X(\Omega_\psi,T,\mu_\psi)\Bigr|_{\eta_\psi =\bar{\eta}_\psi},
\nonumber \\
\label{pmsetaphi}
\\ \bar{\eta}_\psi^2&=&\frac{\lambda_\psi}{3}\tilde{\psi}^2+\frac{\lambda}{2}
\tilde{\phi}^2 \nonumber \\ &+&\frac{2\lambda_\psi }{3}
X(\Omega_\psi,T,\mu_\psi)\Bigr|_{\eta_\psi =\bar{\eta}_\psi}+ \lambda
X(\Omega_\phi,T,\mu_\phi)\Bigr|_{\eta_\phi=\bar{\eta}_\phi },
\nonumber \\
\label{pmsetapsi}
\end{eqnarray}
which are to be solved together with the ones defining the background
field values  $\tilde{\phi}$ and $\tilde{\psi}$, obtained from,
\begin{equation} \label{effmass}
\frac{\partial V_{\rm eff, R}}{\partial \phi_0}\Bigr|_{\phi_0=\tilde
  \phi, \psi_0=\tilde \psi}=0, \;\;\;\; \frac{\partial V_{\rm eff,
    R}}{\partial \psi_0}\Bigr|_{\phi_0=\tilde \phi, \psi_0=\tilde
  \psi}=0,
\end{equation}
which give, respectively, the expressions,
\begin{eqnarray}
&&\tilde{\phi}\left[ m_{\phi }^{2}-\mu_\phi^2+\frac{\lambda_{\phi
    }}{6}\tilde{\phi}^{2} +\frac{\lambda}{2} \tilde{\psi}^{2} \right.
    \nonumber \\ && \left. +\frac{2\lambda_{\phi
    }}{3}X(\Omega_\phi,T,\mu_\phi)\Bigr|_{\eta_\phi=\bar{\eta}_\phi}
    +\lambda X(\Omega_\psi,T,\mu_\psi)\Bigr|_{\eta_\psi
      =\bar{\eta}_\psi}  \right] =0,  \nonumber \\
\label{VEVphi0}
\\ &&\tilde{\psi}\left[ m_{\psi }^{2} - \mu_\psi^2
  +\frac{\lambda_{\psi }}{6}\tilde{\psi}^{2} +\frac{\lambda}{2}
  \tilde{\phi}^{2} \right.  \nonumber \\ &&
  \left. +\frac{2\lambda_{\psi
  }}{3}X(\Omega_\psi,T,\mu_\psi)\Bigr|_{\eta_\psi=\bar{\eta}_\psi}
  +\lambda X(\Omega_\phi,T,\mu_\phi)\Bigr|_{\eta_\phi
    =\bar{\eta}_\phi}  \right] =0.  \nonumber \\
\label{VEVpsi0}
\end{eqnarray}
Equations~(\ref{VEVphi0}) and (\ref{VEVpsi0}) have the trivial
solutions, $\tilde \phi=\tilde \psi=0$, while  the coupled gap
equations valid when  $\tilde \phi\neq 0$ and $\tilde \psi\neq 0$ are
given, respectively, by
\begin{eqnarray}
&&m_{\phi }^{2}-\mu_\phi^2+\frac{\lambda_{\phi }}{6}\tilde{\phi}^{2}
  +\frac{\lambda}{2} \tilde{\psi}^{2} +\frac{2\lambda_{\phi
  }}{3}X(\Omega_\phi,T,\mu_\phi)\Bigr|_{\eta_\phi=\bar{\eta}_\phi}
  \nonumber \\ && +\lambda X(\Omega_\psi,T,\mu_\psi)\Bigr|_{\eta_\psi
    =\bar{\eta}_\psi}  =0, 
\label{VEVphi}
\\ && m_{\psi }^{2} - \mu_\psi^2 +\frac{\lambda_{\psi
}}{6}\tilde{\psi}^{2} +\frac{\lambda}{2} \tilde{\phi}^{2}
+\frac{2\lambda_{\psi
}}{3}X(\Omega_\psi,T,\mu_\psi)\Bigr|_{\eta_\psi=\bar{\eta}_\psi}
\nonumber \\ &&  +\lambda X(\Omega_\phi,T,\mu_\phi)\Bigr|_{\eta_\phi
  =\bar{\eta}_\phi}  =0. 
\label{VEVpsi}
\end{eqnarray}

Note that by combining Eqs.~(\ref{pmsetaphi}) and (\ref{pmsetapsi})
with Eqs.~(\ref{VEVphi}) and (\ref{VEVpsi}), we obtain
\begin{equation}
\tilde{\phi}^{2} = \frac{3}{\lambda_\phi} \left( m_\phi^2 - \mu_\phi^2
+ \bar{\eta}_\phi^2 \right),
\label{phi2eta}
\end{equation}
and
\begin{equation}
\tilde{\psi}^{2} = \frac{3}{\lambda_\psi} \left( m_\psi^2 - \mu_\psi^2
+ \bar{\eta}_\psi^2 \right),
\label{psi2eta}
\end{equation}
which gives that at the critical point for, e.g., in the
$\phi$-direction, $\tilde{\phi}(T_{c,\phi},\mu_{c,\phi})=0$, the
critical chemical potential gets uniquely fixed by 
\begin{equation}
\mu_{\phi,c}^2= m_\phi^2 + \bar{\eta}_\phi^2(T_{c,\phi},\mu_{c,\phi}),
\label{muphic}
\end{equation}
and, equivalently, at the critical point in the $\psi$-direction,
$\tilde{\psi}(T_{c,\psi},\mu_{c,\psi})=0$, it leads to
\begin{equation}
\mu_{\psi,c}^2= m_\psi^2 + \bar{\eta}_\psi^2(T_{c,\psi},\mu_{c,\psi}).
\label{mupsic}
\end{equation}
Equations (\ref{muphic}) and (\ref{mupsic}) generalize to the present
problem the usual condition for Bose-Einstein condensation (BEC) for
an ideal Bose gas~\cite{Kapusta:2006pm}, $|\mu| = |m|$ at the critical
temperature for BEC. Note that the contribution from the interactions
to the BEC transition point enters implicitly in the OPT functions,
$\bar{\eta}_\phi^2$ and  $\bar{\eta}_\psi^2$, in Eqs.~(\ref{muphic})
and (\ref{mupsic}).

\subsection{The densities}

{}From the effective thermodynamical  potential, we can compute the
pressure,
\begin{equation}
P(\mu_\phi,\mu_\phi,T) = - V_{\rm eff,R}^{(\delta)}\Bigr|_{
  \bar{\eta}_\phi,\bar{\eta}_\psi,\tilde{\phi},\tilde{\psi} },
\label{pressure}
\end{equation}
which is evaluated at the PMS values $\eta_\phi=\bar{\eta}_\phi$ and
$\eta_\psi=\bar{\eta}_\psi$ and at the VEV values
$\phi_0=\tilde{\phi}$ and $\psi_0=\tilde{\psi}$.  Given the pressure,
the densities are evaluated as
\begin{eqnarray}
&& n_\phi = \frac{\partial P(\mu_\phi,\mu_\phi,T) }{\partial
    \mu_\phi},
\label{nphi0}
\\ && n_\psi = \frac{\partial P(\mu_\phi,\mu_\phi,T) }{\partial
  \mu_\psi}.
\label{npsi0}
\end{eqnarray}
Then, making use of Eq.~(\ref{VeffR}) in combination with the PMS
equation (\ref{pms}), we find
\begin{eqnarray}
n_\phi &=&\mu_\phi \tilde{\phi}^2 - \frac{\partial
  Y(\Omega_\phi,\mu_\phi,T) }{\partial
  \mu_\phi}\Bigr|_{\bar{\eta}_\phi},
\label{nphi}
\\ n_\psi &=& \mu_\psi \tilde{\psi}^2 - \frac{\partial
  Y(\Omega_\psi,\mu_\psi,T) }{\partial
  \mu_\psi}\Bigr|_{\bar{\eta}_\psi}.
\label{npsi}
\end{eqnarray}
Note that the above equations for $ n_\phi$ and $ n_\psi$ are to be
solved simultaneously with those for the PMS,
Eqs.~(\ref{pmsetaphi})and (\ref{pmsetapsi}), together with those for
the background fields,  Eqs.~(\ref{VEVphi0}) and (\ref{VEVpsi0}).

\section{ISB and SNR in a thermal and dense medium: results}
\label{results}

When the square masses $m_\phi^2$ and $m_\psi^2$ are both positive in
the tree-level potential Eq.~(\ref{potential}) and under appropriate
choice of coupling constants satisfying the boundness condition for
the potential, we have the possibility of having ISB in one of the
directions at high temperatures, while the other field remains in the
symmetric phase.  On the other hand, when the square masses $m_\phi^2$
and $m_\psi^2$ are both negative in the tree-level potential
Eq.~(\ref{potential}), we have the possibility of having SNR for one
of the fields at high temperatures, while the other one will suffer
the usual symmetry restoration at some critical temperature $T_{c}$.

\begin{center}
\begin{figure}[!htb]
\includegraphics[width=7.5cm]{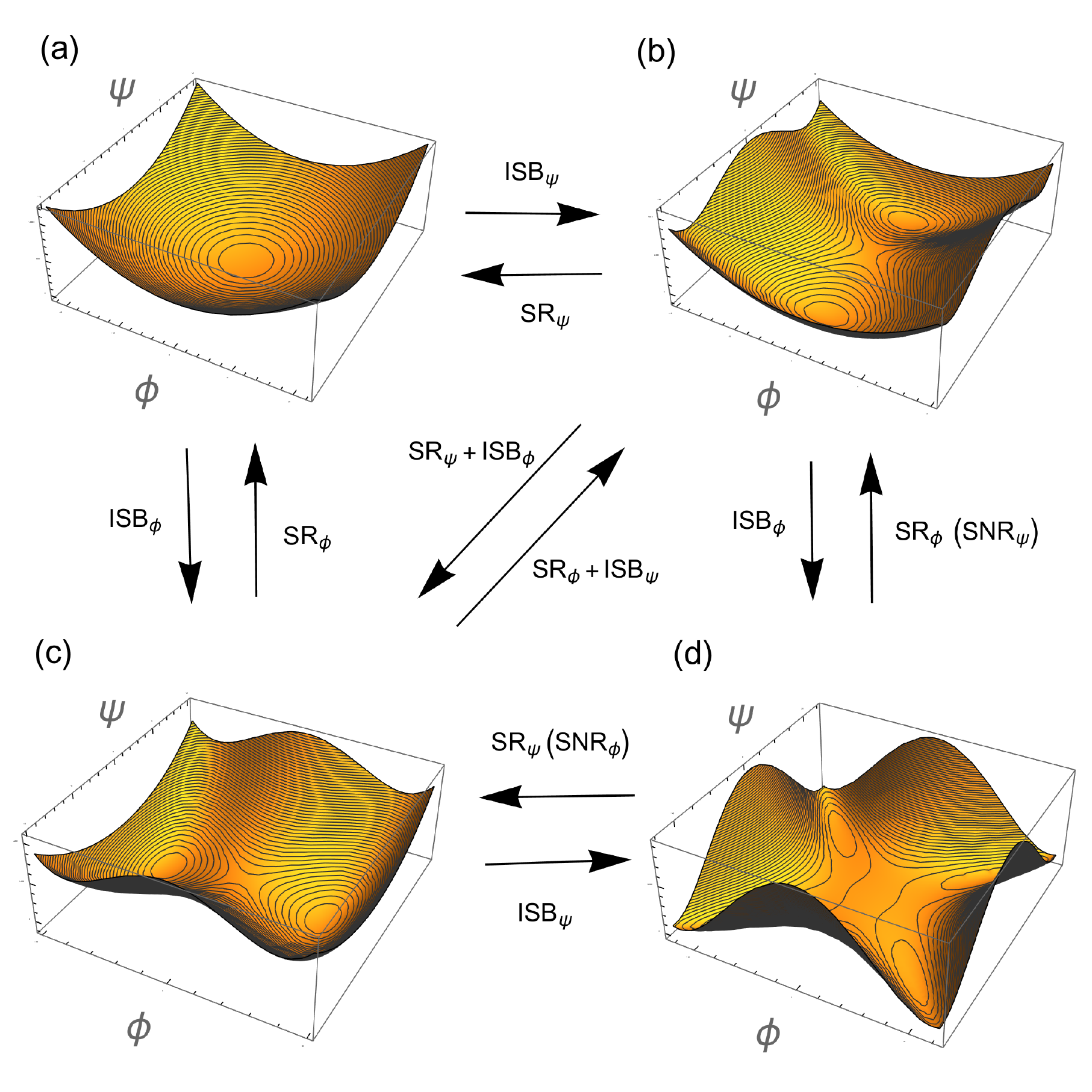}
\caption{The different phases allowed by the system and the possible
  directions for phase transitions when varying $T$ and/or chemical
  potentials $\mu_\phi$ and $\mu_\psi$, or densities.}
\label{fig2}
\end{figure}
\end{center}

{}Figure~\ref{fig2} illustrates the different phases in which the
system might display depending on the choices  made for the model
parameters, temperature and density. Let us describe the four cases
illustrated in {}Fig.~\ref{fig2}. Case (a): the system is in the
symmetric state with respect to the two fields, $\langle \phi \rangle
=\langle\psi\rangle=0$. Case (b): the system is in a state with
symmetry breaking in the direction of $\phi$, $\langle \phi
\rangle\neq 0$, and symmetry restored  in the direction of $\psi$,
$\langle \psi \rangle= 0$. Case (c): the system is in a state with
symmetry breaking in the direction of $\psi$, $\langle \psi
\rangle\neq 0$, and symmetry restored  in the direction of $\phi$,
$\langle \phi \rangle= 0$. Case (d): the system is in the symmetry
broken  state with respect to the two fields,  $\langle \phi \rangle
\neq  0 $ and $\langle\psi\rangle\neq 0$. The arrows indicate the
possible transitions that the system might experience when changing
$T$ and/or $\mu_\phi$ and $\mu_\psi$. 

In the results below, we will analyze these different cases. We will
first analyze the situation where ISB becomes  possible at high
temperatures, and then the situation where SNR becomes viable.  {}For
convenience and without loss of generality, we will assume $m_\phi^2 =
m_\psi^2 = m^2$ and $\mu_\phi = \mu_\psi=\mu$.  All quantities are
normalized by the regularization scale $M$.

\subsection{The ISB case: $m_\phi^2>0$ and $m_\psi^2>0$ }

{}For our numerical results we will consider for illustration purposes
the base parameters values for the couplings: $\lambda_\phi=0.018$,
$\lambda_\psi=0.6$ and $\lambda=-0.03$. 
Note that though $\lambda < 0$, we have that 
$\lambda_\phi \lambda_\psi = 0.0108 > 9 \lambda^2 =0.0081$, thus,
this choice of couplings
satisfies the boundness condition for the potential. 
In other words, the tree potential is safely inside the
stable region.
Other choices can
be made but the results are qualitatively similar under the conditions
considered here.  Thus, by considering $m_\phi^2 = m_\psi^2 = m^2>0$
and the set of coupling parameters given above,  ISB can be shown to
happen in the direction of $\phi$ at high temperatures, while $\psi$
remains in the symmetric phase. This is what simple perturbation
theory at high temperatures would predict for the two-field coupled
complex scalar model in the absence of chemical
potentials~\cite{Farias:2021ult}.  Note that we can recover the PT
results from the OPT interpolated ETP Eq.~(\ref{VeffR}) by setting
$\eta_\phi =\eta_\psi=0$, which then gives the ETP at first order in
the coupling constants in the first order PT  approximation. In the
OPT case, we do obtain though, quantitative differences when compared
with the results from PT as we are going to illustrate.

\begin{center}
\begin{figure}[!htb]
\includegraphics[width=7.5cm]{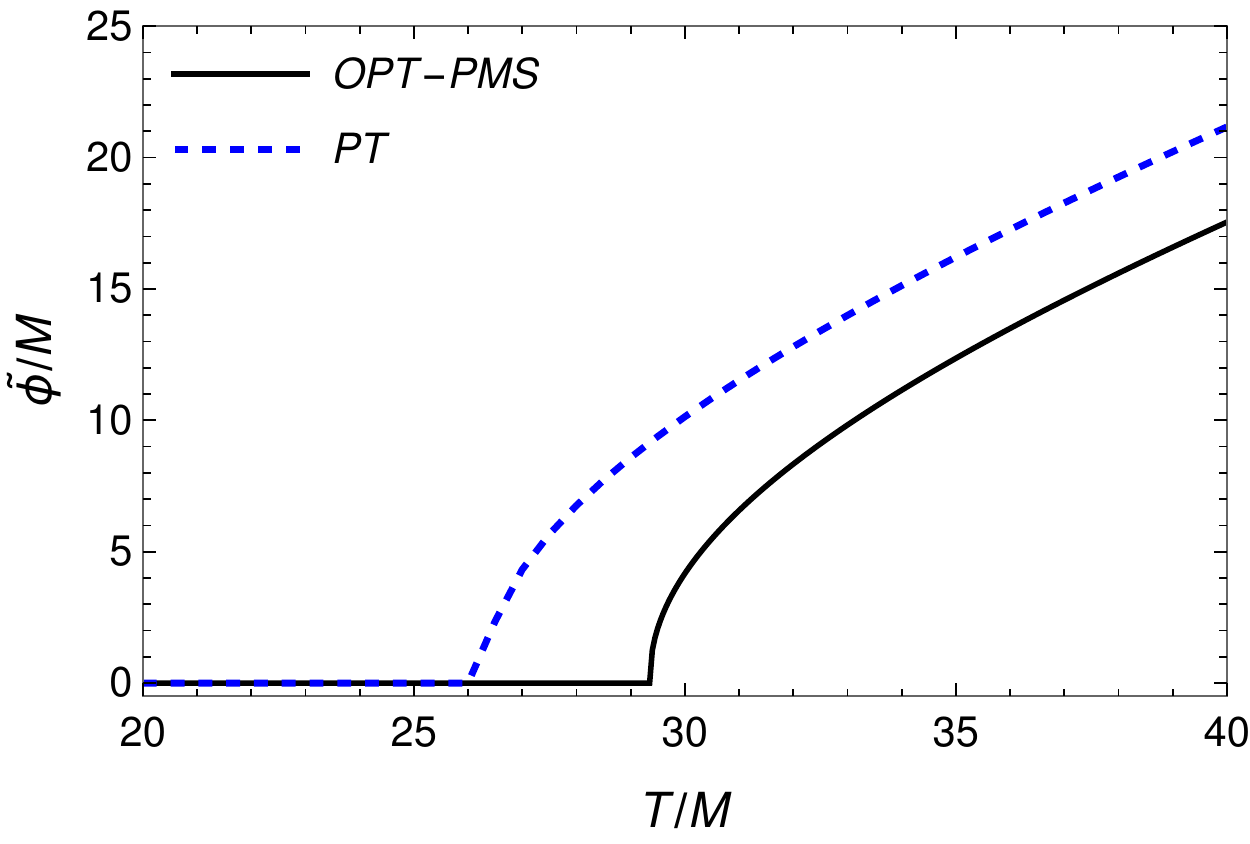}
\caption{The VEV $\tilde \phi$ for OPT and PT as a function of the
  temperature (at $\mu=0$).  The parameters considered are $m_\phi^2 =
  m_\psi^2 = m^2 >0$, $M=m$, $\mu_\phi = \mu_\psi=\mu$,
  $\lambda_\phi=0.018$, $\lambda_\psi=0.6$, and $\lambda=-0.03$.}
\label{fig3}
\end{figure}
\end{center}

\begin{center}
\begin{figure}[!htb]
\includegraphics[width=7.5cm]{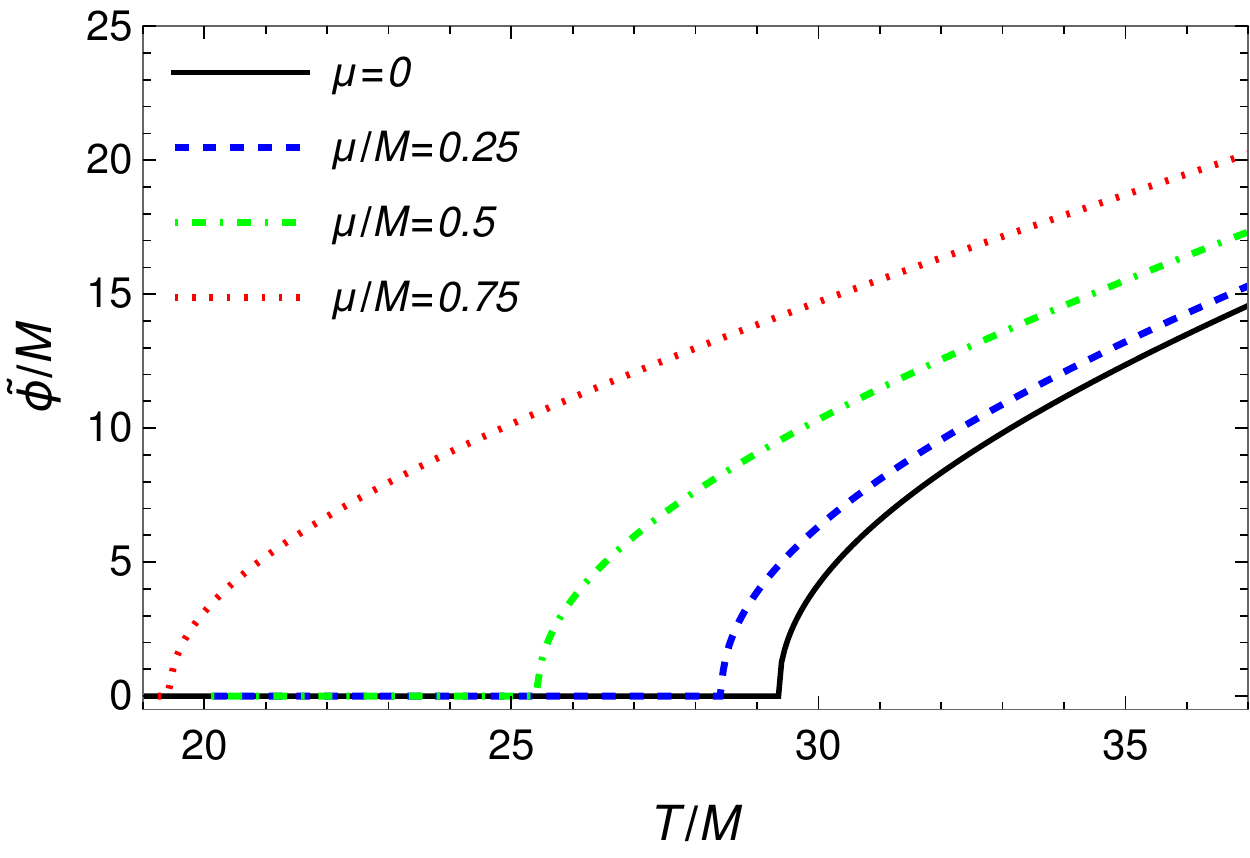}
\caption{The VEV $\tilde \phi$ in the OPT case as a function of the
  temperature and for different values of the chemical potential. The
  model parameters are the same as considered in {}Fig.~\ref{fig3}.}
\label{fig4}
\end{figure}
\end{center}

\begin{center}
\begin{figure}[!htb]
\includegraphics[width=7.5cm]{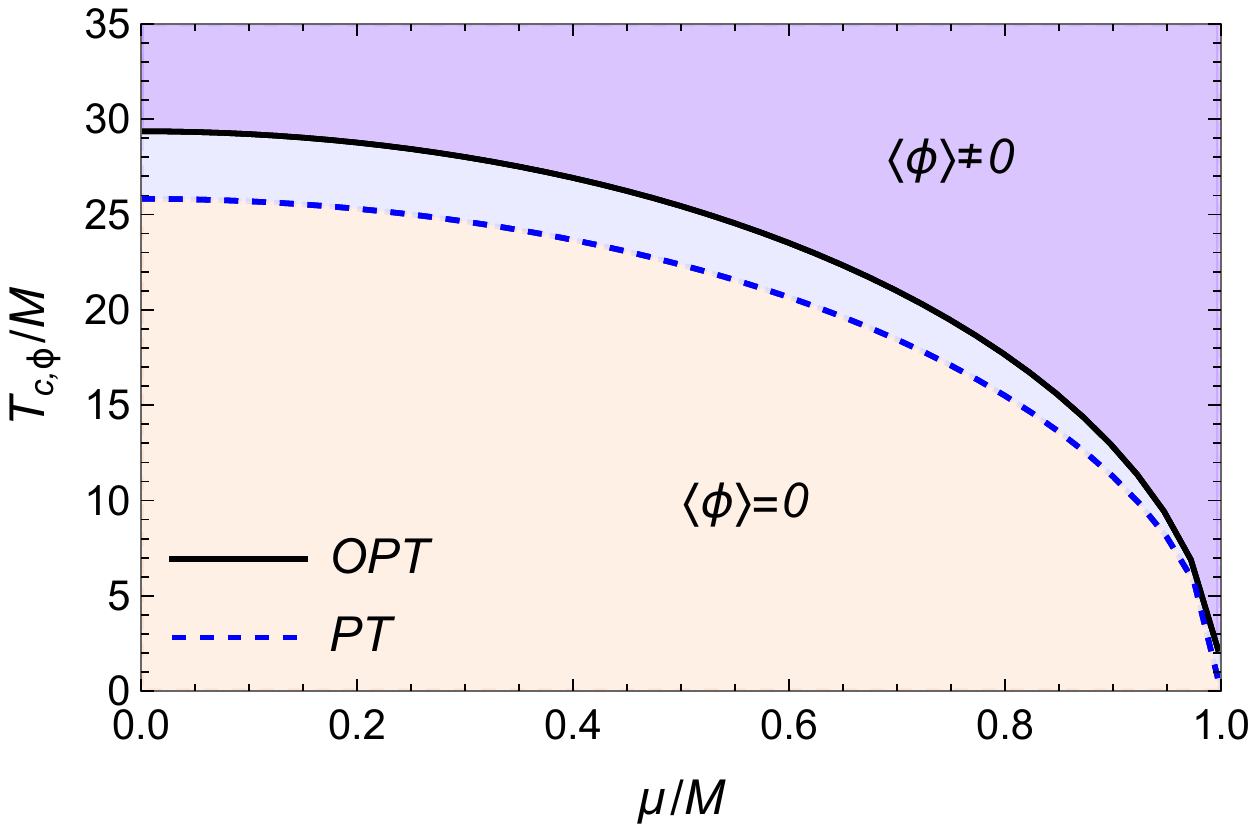}
\caption{The critical temperature for ISB in the $\phi$-field
  direction as a function of the chemical potential.  Here,  both OPT
  and the PT approximation are considered for  comparison.  }
\label{fig5}
\end{figure}
\end{center}

The situation illustrated in the {}Fig.~\ref{fig3} demonstrates ISB in
the direction of the field $\phi$.  
Starting from a symmetry restored
(SR) state at $T=0,\, \mu_\phi=\mu_\psi=0$, the $\phi$ field
eventually acquires a nonvanishing VEV at a critical temperature.
Both OPT and PT are compared. One notices that the OPT tends to
produce a higher critical temperature than in the PT approximation
case. Given the parameters considered, the field $\psi$ remains in the
SR phase. The effect of a finite chemical potential on the behavior
for the VEV of the $\phi$ field is illustrated in the
{}Fig.~\ref{fig4}. 
Note that for our choice of parameters and for $\mu$ held constant
at the values considered, the field $\psi$ remains always 
with a vanishing background expectation value, $\tilde \psi=0$, hence, we do not show it
in {}Figs.~\ref{fig3} and \ref{fig4}.
In {}Fig.~\ref{fig4}, we restrict only to show the OPT case, since for PT
the results are similar, with the same trend as seen in
{}Fig.~\ref{fig3}, leading to smaller critical temperatures when
compared to the OPT.  {}From the results shown in {}Fig.~\ref{fig4},
we see that the chemical potential tends to decrease the critical
temperature for ISB. Thus, the larger is the chemical potential, the
easier is to reach symmetry inversion in the direction of $\phi$,
i.e., the $\phi$ field acquires a VEV $\langle \phi \rangle\neq 0$ at
lower temperatures as  the chemical potential increases.  As the field
changes smoothly through the critical point, the phase transition
associated with ISB here is of second order.  The overall behavior for
the critical temperature for ISB in the direction of $\phi$,
$T_{c,\phi}$, as a function of the chemical potential, is shown in
{}Fig.~\ref{fig5}.

The {}Fig.~\ref{fig5} shows that the higher the chemical potential,
the lower the value of the critical temperature. The presence of
charge is seen here to favor symmetry inversion (ISB), causing it to
happen at a lower critical temperature.

{}From the mass eigenvalues expressions given in Appendix~\ref{appB}, we
can  define the corresponding Higgs and Goldstone effective modes for
$\phi$ and $\psi$ in the context of the OPT.  This can be done by
introducing in the definitions Eq.~(\ref{MHMG}), the thermal and
chemical potential contributions such
that~\cite{Duarte:2011ph,Farias:2021ult}, $M_{H,\phi}^2\to
M_{H,\phi}^2(T,\mu_\phi,\mu_\psi)$, $M_{G,i}^2\to
M_{G,\phi}^2T,\mu_\phi,\mu_\psi)$ and similarly for the $\psi$ field,
where, at first order in the OPT,
\begin{eqnarray}
M_{H,\phi}^2(T,\mu_\phi,\mu_\psi) &=& m_\phi^2-\mu_\phi^2 +
\frac{\lambda_\phi}{2}\tilde{\phi}^2 +\frac{\lambda}{2}\tilde{\psi}^2
\nonumber \\ &+& \frac{2 \lambda_\phi}{3}
X(\Omega_\phi,T,\mu_\phi)\Bigr|_{\eta_\phi=\bar{\eta}_\phi}  \nonumber
\\ &+&  \lambda X(\Omega_\psi,T,\mu_\psi)\Bigr|_{\eta
  _\psi=\bar{\eta}_\psi},
\label{mphi1eff}
\\  M_{G,\phi}^2(T,\mu_\phi,\mu_\psi) &=& m_\phi^2 +
\frac{\lambda_\phi}{6}\tilde{\phi}^2 +\frac{\lambda}{2}\tilde{\psi}^2
\nonumber \\ &+& \frac{2 \lambda_\phi}{3}
X(\Omega_\phi,T,\mu_\phi)\Bigr|_{\eta_\phi=\bar{\eta}_\phi}  \nonumber
\\ &+&  \lambda X(\Omega_\psi,T,\mu_\psi)\Bigr|_{\eta
  _\psi=\bar{\eta}_\psi},
\label{mphi2eff}
\\  M_{H,\psi}^2(T,\mu_\phi,\mu_\psi) &=& m_\psi^2
+\frac{\lambda_\psi}{2}\tilde{\psi}^2 +
\frac{\lambda}{2}\tilde{\phi}^2 \nonumber \\ &+& \frac{2
  \lambda_\psi}{3} X(\Omega_\psi,T,\mu_\psi)\Bigr|_{\eta
  _\psi=\bar{\eta}_\psi}  \nonumber \\ &+& \lambda
X(\Omega_\phi,T,\mu_\phi)\Bigr|_{\eta _\phi=\bar{\eta}_\phi},
\label{mpsi1eff} 
\\  M_{G,\psi}^2(T,\mu_\phi,\mu_\psi) &=& m_\psi^2
+\frac{\lambda_\psi}{6}\tilde{\psi}^2 +
\frac{\lambda}{2}\tilde{\phi}^2 \nonumber \\ &+& \frac{2
  \lambda_\psi}{3} X(\Omega_\psi,T,\mu_\psi)\Bigr|_{\eta
  _\psi=\bar{\eta}_\psi}  \nonumber \\ &+& \lambda
X(\Omega_\phi,T,\mu_\phi)\Bigr|_{\eta _\phi=\bar{\eta}_\phi},
\label{mpsi2eff} 
\end{eqnarray}
where $\tilde{\phi}$ and $\tilde{\psi}$ are obtained from the solution
of Eq.~(\ref{effmass}), while $\bar{\eta}_\phi$ and $\bar{\eta}_\psi$
from the PMS equations (\ref{pmsetaphi}) and (\ref{pmsetapsi}).  Note
that the Higgs modes must remain positive in both symmetric and broken
phases, while vanishing at the critical point for phase
transition. The Goldstone modes, on the other hand, must remain
massless in the broken phases, according to the Goldstone theorem
concerning symmetry breaking of a continuous symmetry, while in the
symmetric phase follows the Higgs modes.

\begin{center}
\begin{figure}[!htb]
\subfigure[]{\includegraphics[width=7.5cm]{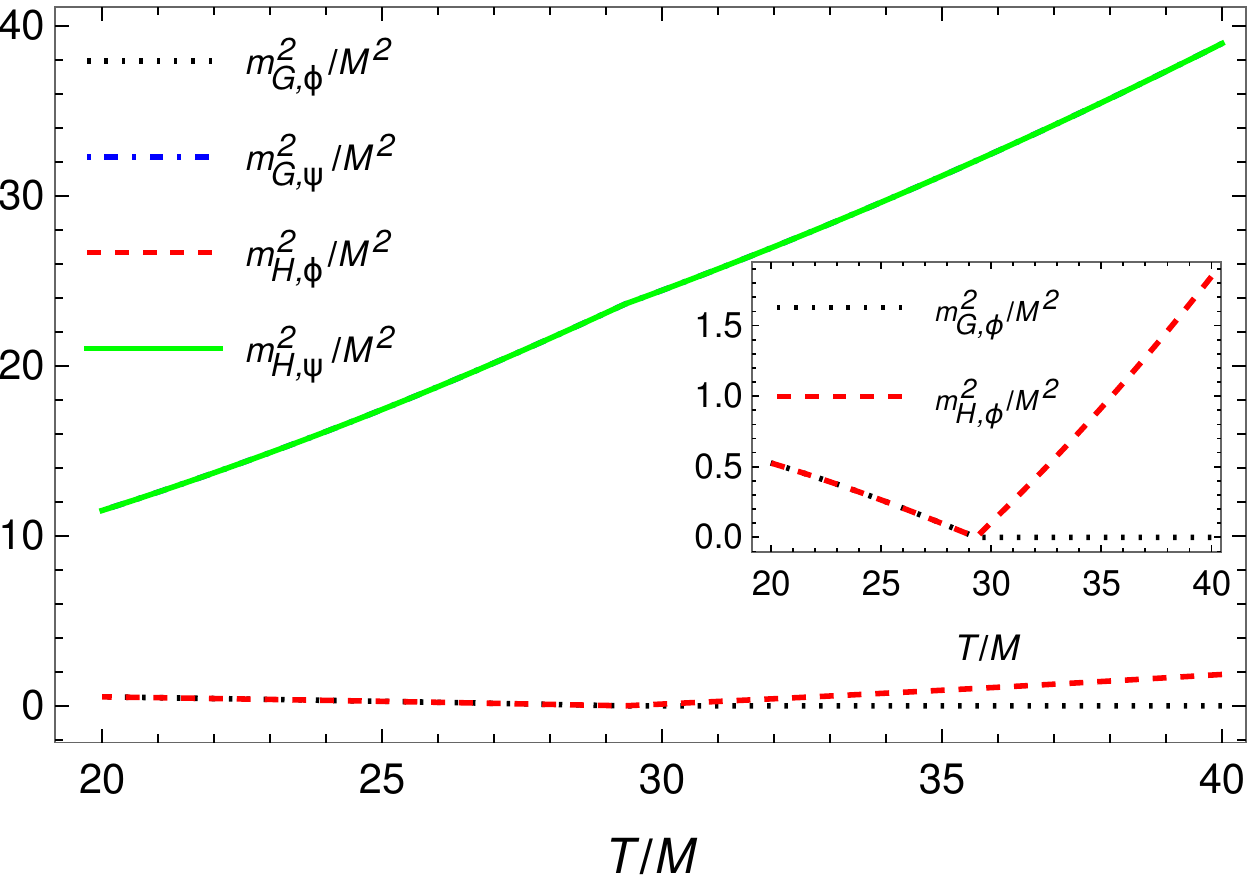}}
\subfigure[]{\includegraphics[width=7.5cm]{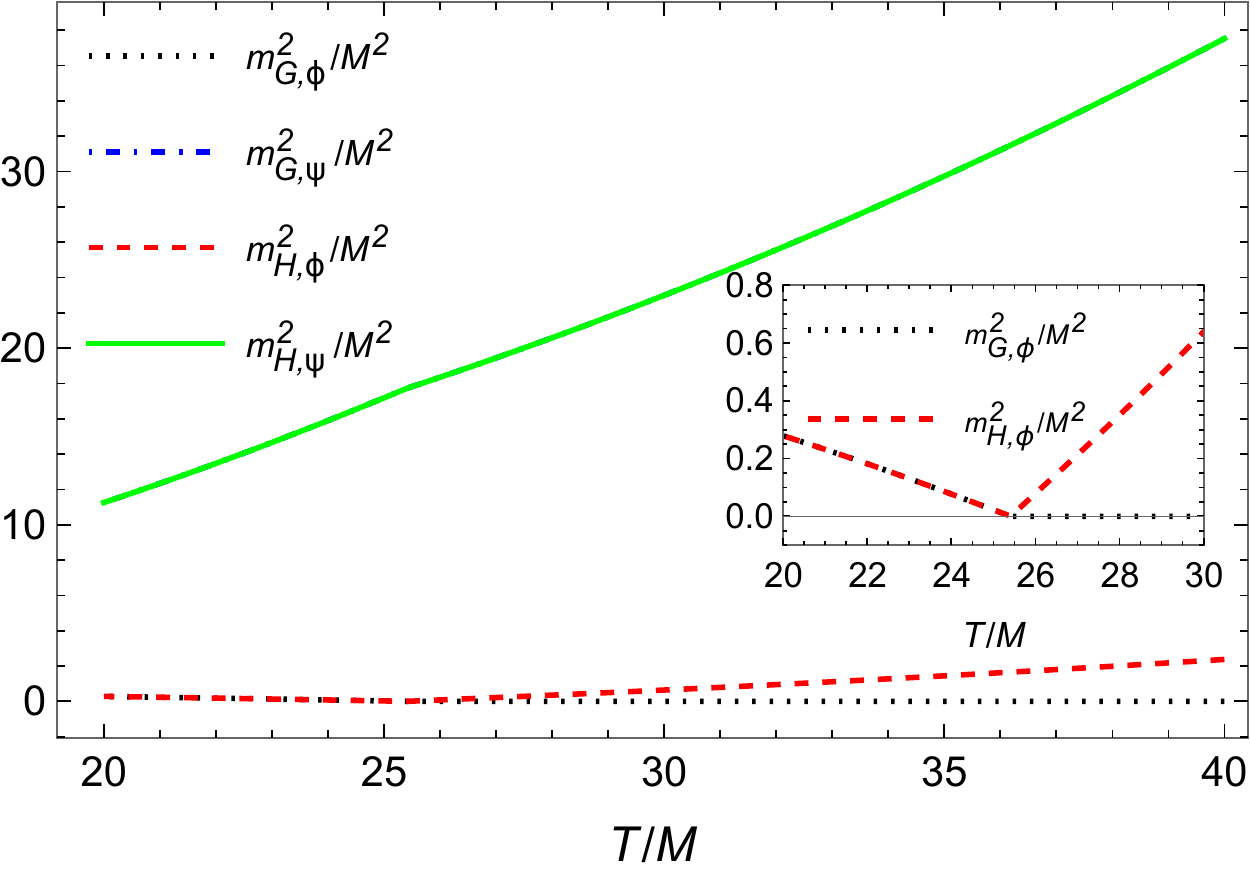}}
\caption{(a) Higgs and Goldstone modes for the fields when
  $\mu_\phi=\mu_\psi=0$.  (b) Higgs and Goldstone modes for the fields
  when $\mu_\phi=\mu_\psi=0.5M$.  In both cases, the curves for
  $m_{H,\psi}$ and $m_{G,\psi}$ since the $\psi$ field remains in  the
  symmetric phase, $\tilde \psi=0$. The model parameters are the same
  as considered in {}Fig.~\ref{fig3}.}
\label{fig6}
\end{figure}
\end{center}

Substituting Eqs.~(\ref{mphi1eff})-(\ref{mpsi2eff}) in the mass
eigenvalue equation (\ref{MHphi})-(\ref{MGpsi}), we obtain the
corresponding ones at finite temperature and chemical potential,
$\mathcal{M}_1^2 \to m_{H,\phi}^2,\; \mathcal{M}_2^2 \to
m_{G,\phi}^2,\; \mathcal{M}_3^2 \to m_{H,\psi}^2,\; \mathcal{M}_4^2
\to m_{G,\psi}^2$. These Higgs and Goldstone modes for each of the
fields are plotted in {}Fig.~\ref{fig6} for the cases of vanishing
chemical potentials (panel a) and also for nonvanishing chemical
potential (panel b).  In both cases we see that the Goldstone theorem
is correctly reproduced. The fact that the OPT satisfies the Goldstone
theorem has also been seen in previous
applications~\cite{Duarte:2011ph,Farias:2021ult}.

\subsection{The SNR case: $m_\phi^2<0$ and $m_\psi^2<0$ }

\begin{center}
\begin{figure}[!htb]
\subfigure[]{\includegraphics[width=7.5cm]{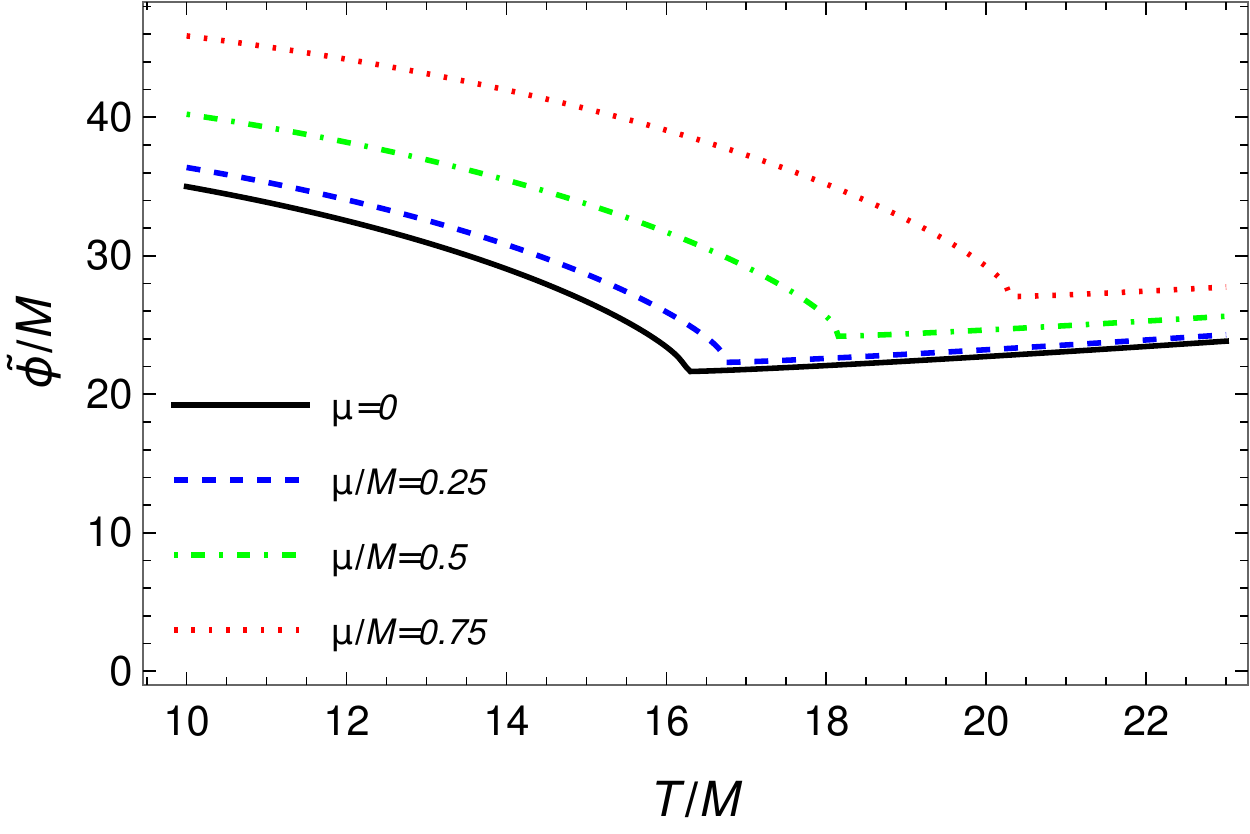}}
\subfigure[]{\includegraphics[width=7.5cm]{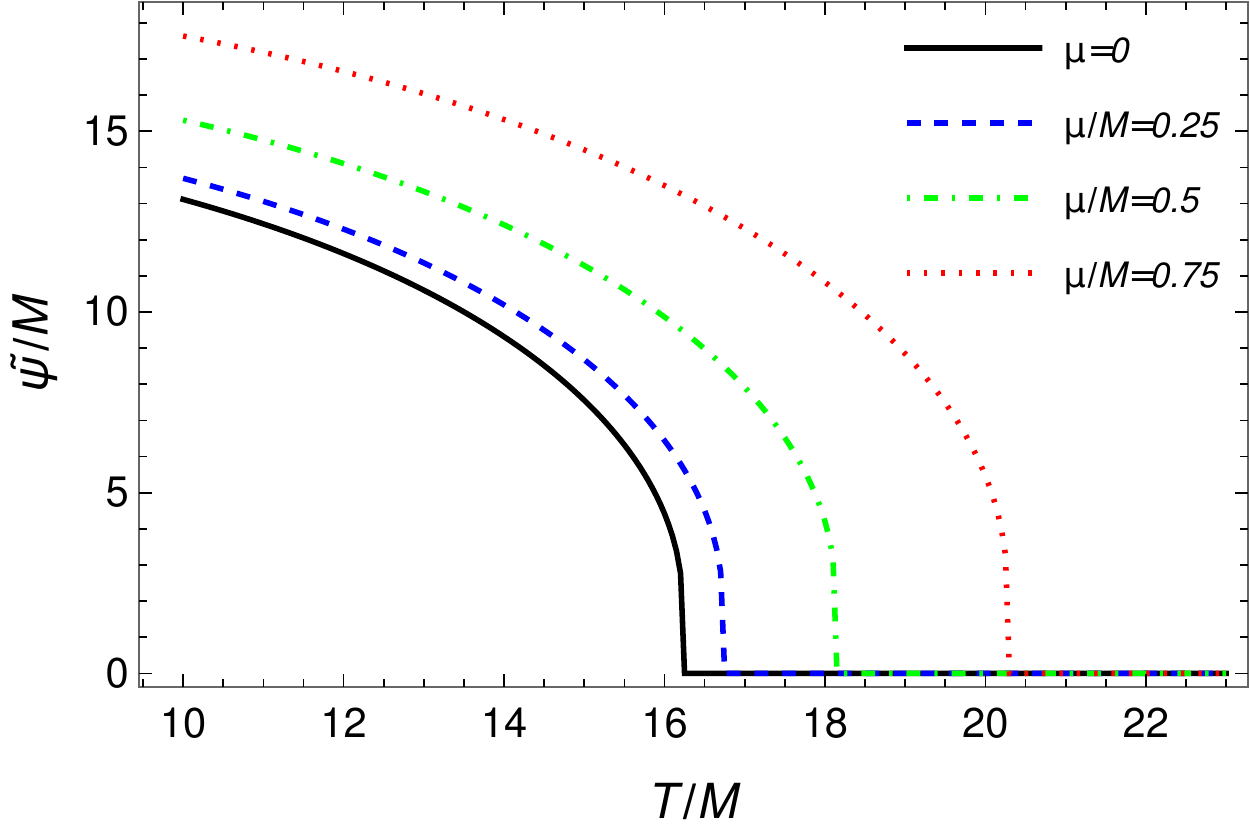}}
\caption{The VEVs $\tilde \phi$ and $\tilde \psi$ for OPT as a
  function of the temperature and for different values of the chemical
  potential. The parameters considered are $m_\phi^2 = m_\psi^2 = -m^2
  < 0$, $M=m$, $\mu_\phi = \mu_\psi=\mu$, $\lambda_\phi=0.018$,
  $\lambda_\psi=0.6$, and $\lambda=-0.03$.}
\label{fig7}
\end{figure}
\end{center}

\begin{center}
\begin{figure}[!htb]
\subfigure[]{\includegraphics[width=7.5cm]{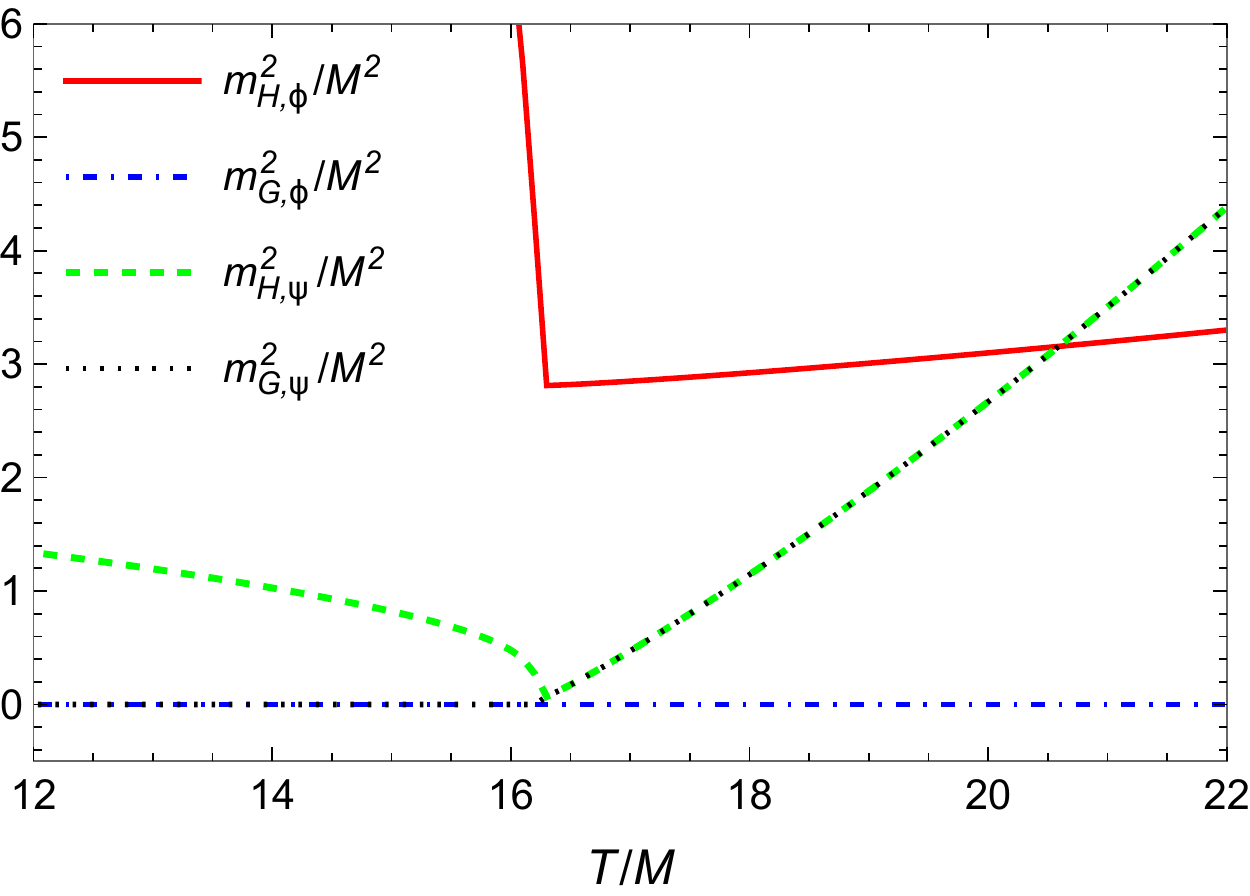}}
\subfigure[]{\includegraphics[width=7.5cm]{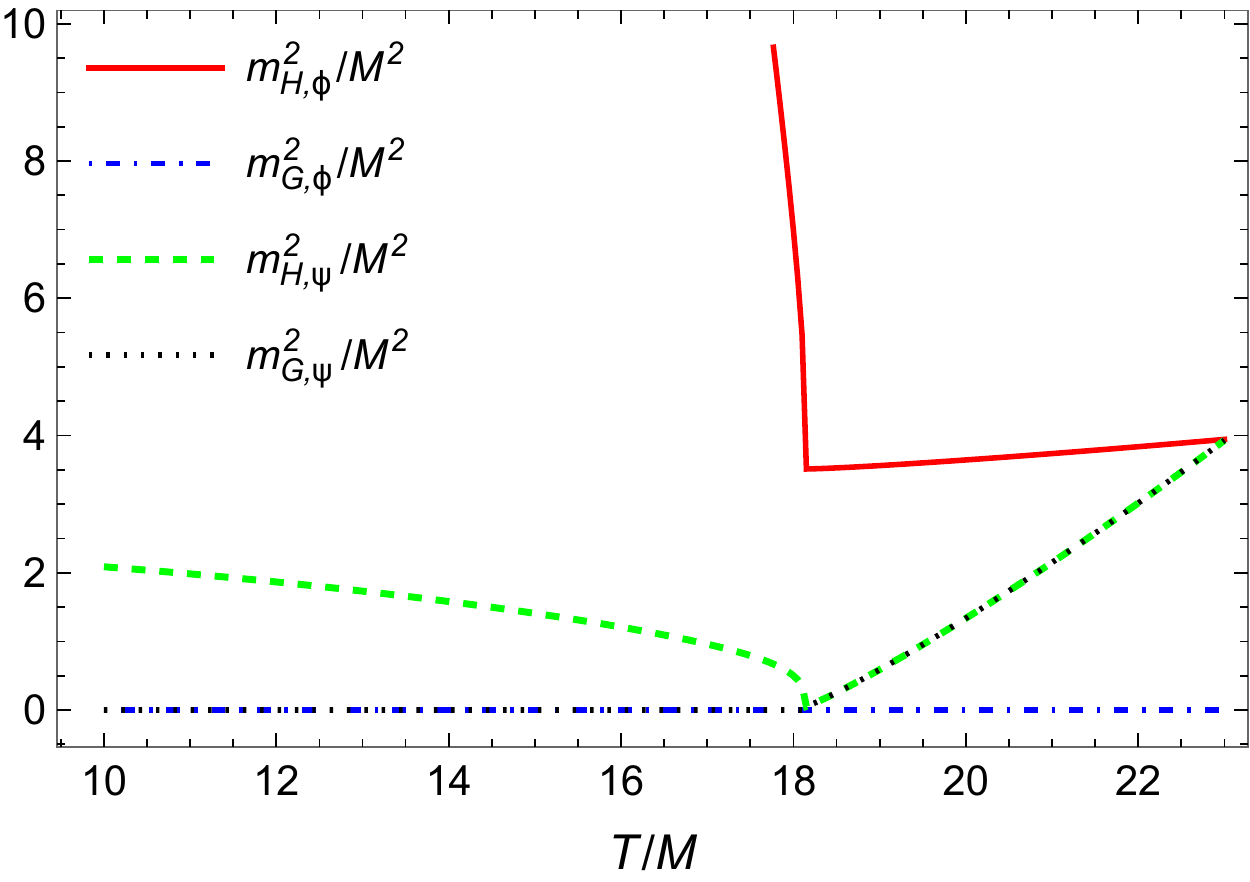}}
\caption{(a) Higgs and Goldstone modes for the fields when
  $\mu_\phi=\mu_\psi=0$.  (b) Higgs and Goldstone modes for the fields
  when $\mu_\phi=\mu_\psi=0.5M$.  In both cases, the curves for
  $m_{H,\psi}$ and $m_{G,\psi}$ since the $\psi$ field remains in  the
  symmetric phase, $\tilde \psi=0$. The model parameters are the same
  as considered in {}Fig.~\ref{fig7}.}
\label{fig8}
\end{figure}
\end{center}

In the previous subsection, we have investigated the two coupled
complex scalar field at finite temperature and chemical potential with
respect to symmetry inversion (ISB).  Let us now study the model for
symmetry persistence at high temperatures, i.e., we will study the
case for SNR in one of the field directions. We continue to use the
same set of bare coupling constants as before for clarity, but now we
consider that the symmetries for both the fields, in the vacuum, are
both broken, $m_\phi^2<0$ and $m_\psi^2<0$, such that both fields have
a nonvanishing VEV at $T=0$ and $\mu_\phi=\mu_\psi=0$ initially.  In
this case, we expect SNR in the direction of $\phi$, while $\psi$
should suffer the usual symmetry restoration (SR) at high
temperatures. This is illustrated in {}Fig.~\ref{fig7}.

In {}Fig.~\ref{fig8} we show the Higgs and Goldstone modes for the
fields when $\mu_\phi=\mu_\psi=0$ (panel a) and when
$\mu_\phi=\mu_\psi=0.5M$ (panel b) for the present case of SNR in the
direction of $\phi$ and SR in the direction of $\psi$. As in the case
studied in the previous subsection, we also see here that the
Goldstone theorem is correctly reproduced.

\subsection{OPT results at finite temperature and density}

Let us now turn our attention of how a finite density affects the
results. In the previous two subsections we were interested in the
effect of a finite chemical potential in the transition patterns
displayed by the two coupled complex scalar field system. Here, we are
interested in investigating the same effects but now at finite
densities. This is mostly motivated by the seminal work performed in
Refs.~\cite{Bernstein:1990kf,Benson:1991nj}. In particular, in
Ref.~\cite{Benson:1991nj} it was shown that finite density effects were
already able to not only delay symmetry restoration at finite
temperature, but also to promote symmetry nonrestoration for a
sufficiently large charge (density).  The novelty result was that both
ISB and SNR could be possible already for a one-field model case.
Here, we are interested in studying how the finite density effects
will further affect the phase transition pattern when considering the
case of more than one coupled complex scalar field. In order to
facilitate the comparison with Ref.~\cite{Benson:1991nj}, we will also
assume densities for the fields such that the ratio of number density
to entropy density is kept fixed. The motivation here, as also in
Ref.~\cite{Benson:1991nj}, stems from the fact that in an adiabatic
expansion, with the entropy remaining constant, the ratio of charge
density over entropy density remains constant. This is like the
expected situation in the case of the expanding Universe in the
radiation dominated phase when in the absence of entropy production
processes. Thus, we assume from now on that the charge density $n_i$
of the fields over entropy energy density remains constant, $n_i/s =
{\rm constant}$.  Since in the ultrarelativistic case, $s\propto T^3$
and $n_i \propto T^3$, we take~\cite{Benson:1991nj}

\begin{equation} 
n_i = \tau_i T^3, 
\label{ratio-ns}
\end{equation}

\noindent
where $\tau_i$ is the proportionality constant\footnote{Note that in
  Ref.~\cite{Benson:1991nj} the proportionality constant was denoted
  by $\eta$. To not confuse with the OPT usual parameter notation, we
  use here instead $\tau_i$ as the proportionality constant. }.  {}For
simplicity, we will also assume $\tau_\phi=\tau_\psi=\tau$. The case
of asymmetries in the charge densities can be implemented without
difficult, which can be of interest in the case of systems with large
differences in the parameters (e.g., in the masses, or which can have
large differences in the way both fields might be coupled to
additional radiation fields in the system).

\begin{center}
\begin{figure}[!htb]
\subfigure[]{\includegraphics[width=7.5cm]{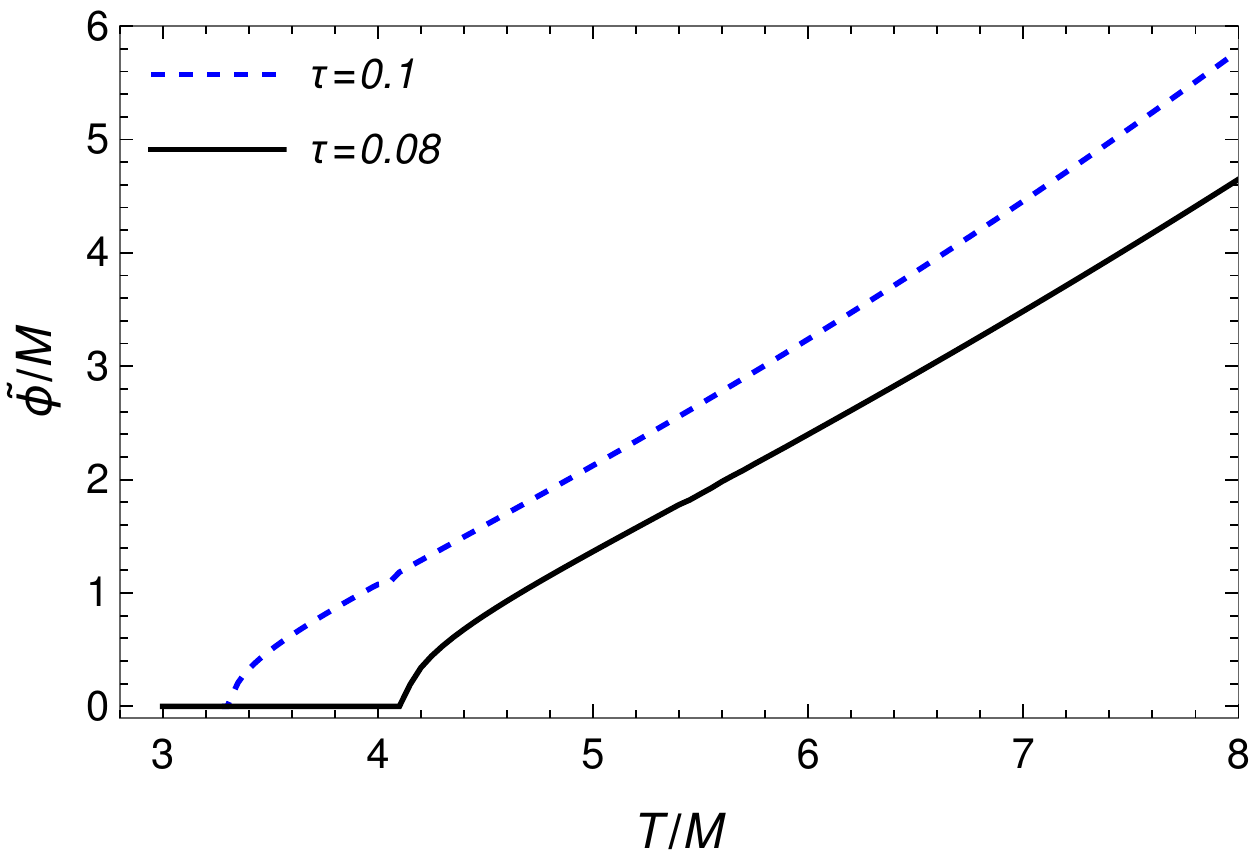}}
\subfigure[]{\includegraphics[width=7.5cm]{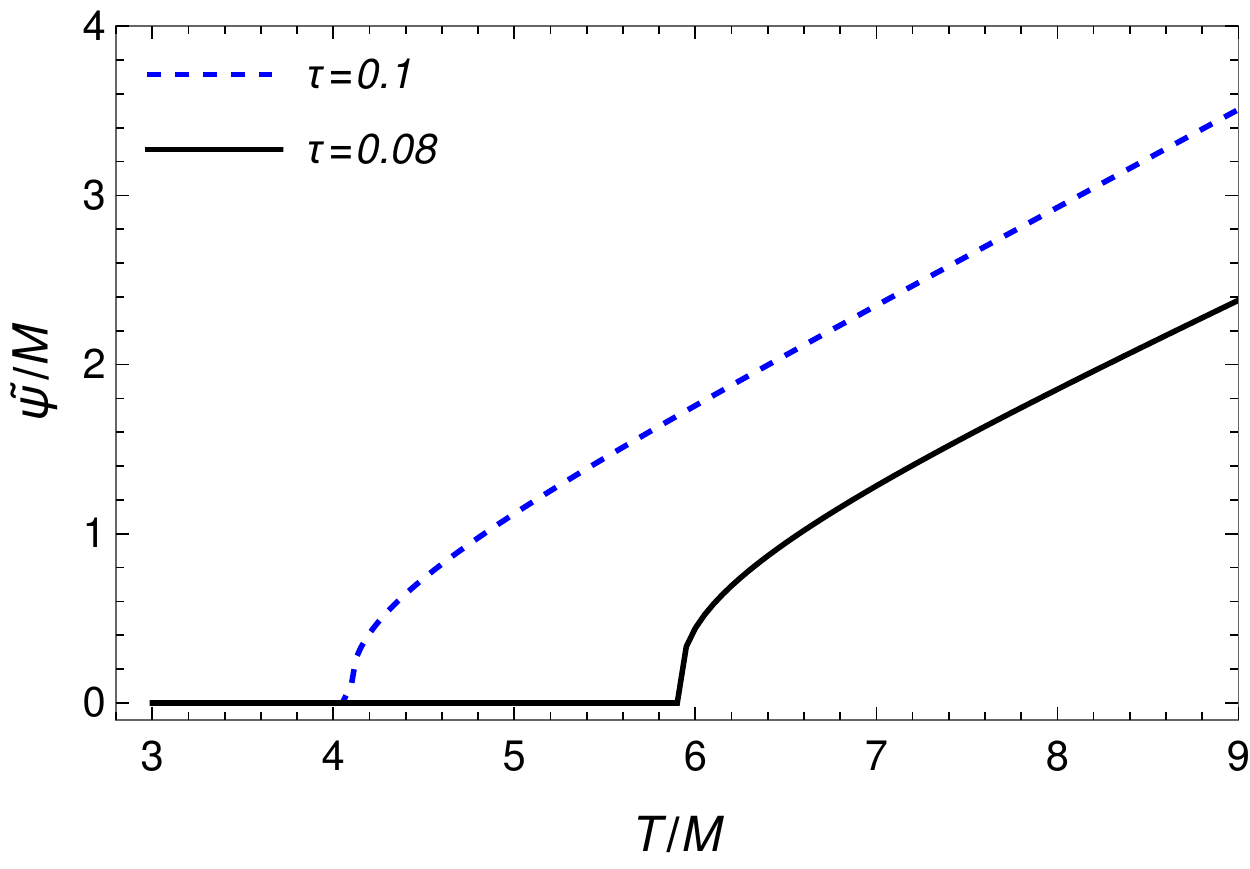}}
\caption{The behavior of the VEV of the fields, $\tilde{\phi}$ (panel
  a) and $\tilde{\psi}$ (panel b), as a function of temperature and
  for two values for the constant parameter $\tau_i$ in the charge density 
expression Eq.~(\ref{ratio-ns}). The parameters are such that $m_\phi^2 = m_\psi^2 = m^2 >0$,
  $M=m$, $\tau_\phi = \tau_\psi=\tau$, $\lambda_\phi=0.018$,
  $\lambda_\psi=0.6$, and $\lambda=-0.03$.}
\label{fig9}
\end{figure}
\end{center}

By using Eq.~(\ref{ratio-ns}) in conjunction with the equations
defining the densities, Eqs.~(\ref{nphi}) and (\ref{npsi}), together
with the PMS and gap equations, we study the behavior of the fields
expectations values  $\tilde{\phi}$ and $\tilde{\psi}$ as a function
of the temperature and fixed ratio $\tau$.  In {}Fig.~\ref{fig9} it is
shown the VEV of the fields, $\tilde{\phi}$ (panel a) and
$\tilde{\psi}$ (panel b), as a function of temperature. Two
representative values of $\tau$ have been used for illustration. Note
that for the parameters considered, in the absence of finite charge
effects ISB is expected to happen in the direction of the $\phi$ field
(see, e.g., {}Fig.~\ref{fig3}). In the presence of finite charge
densities, we see that two effects appear. {}First, the ISB transition
can happen earlier, i.e., at a lower critical temperature. Second,
there is now also an ISB transition in the $\psi$ field direction,
i.e., $\tilde{\psi}$ can now acquire a nonvanishing value at finite
temperatures and also here we see that the larger is $\tau$, the lower
is the critical temperature for ISB in the direction of
$\psi$. {}Finite charges are then realizing the transition pattern (a)
$\to$ (b) $\to$ (d) shown in {}Fig.~\ref{fig1} in the present case.

\begin{center}
\begin{figure}[!htb]
\subfigure[]{\includegraphics[width=7.5cm]{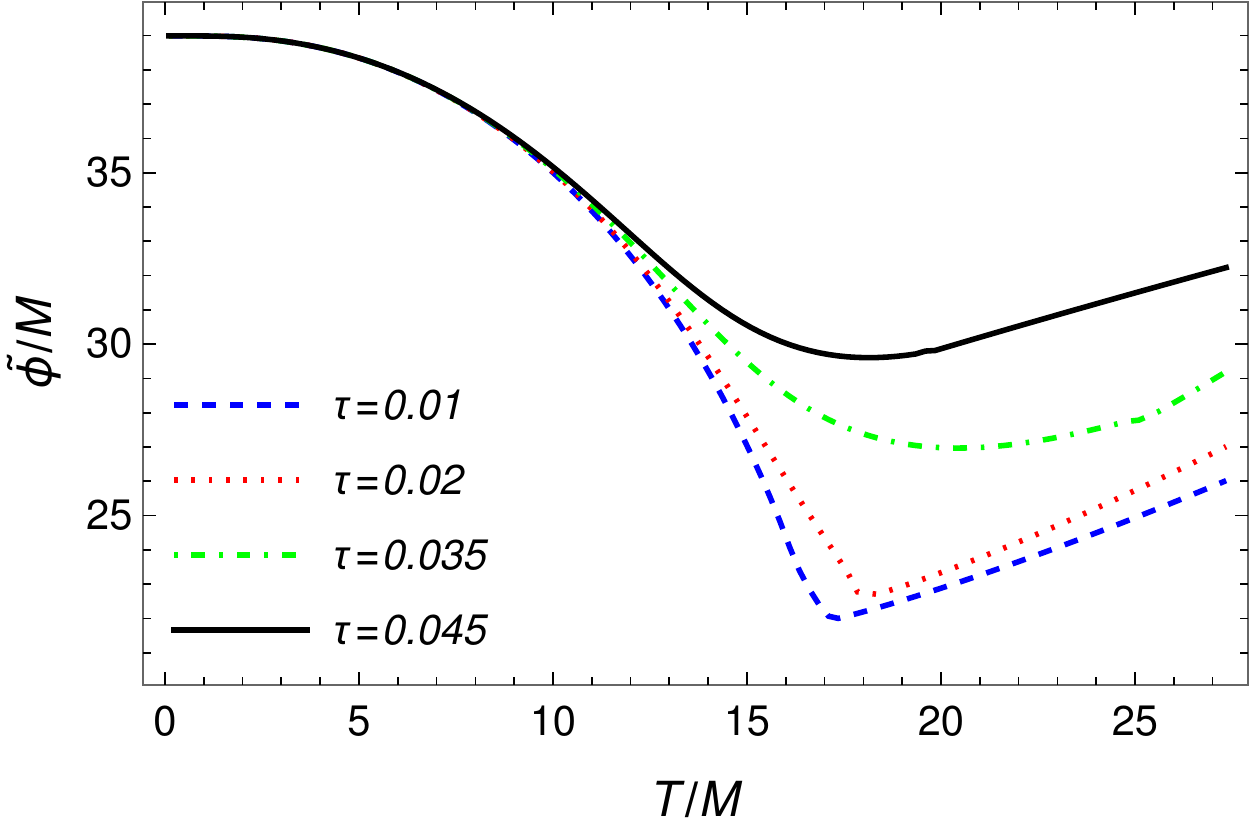}}
\subfigure[]{\includegraphics[width=7.5cm]{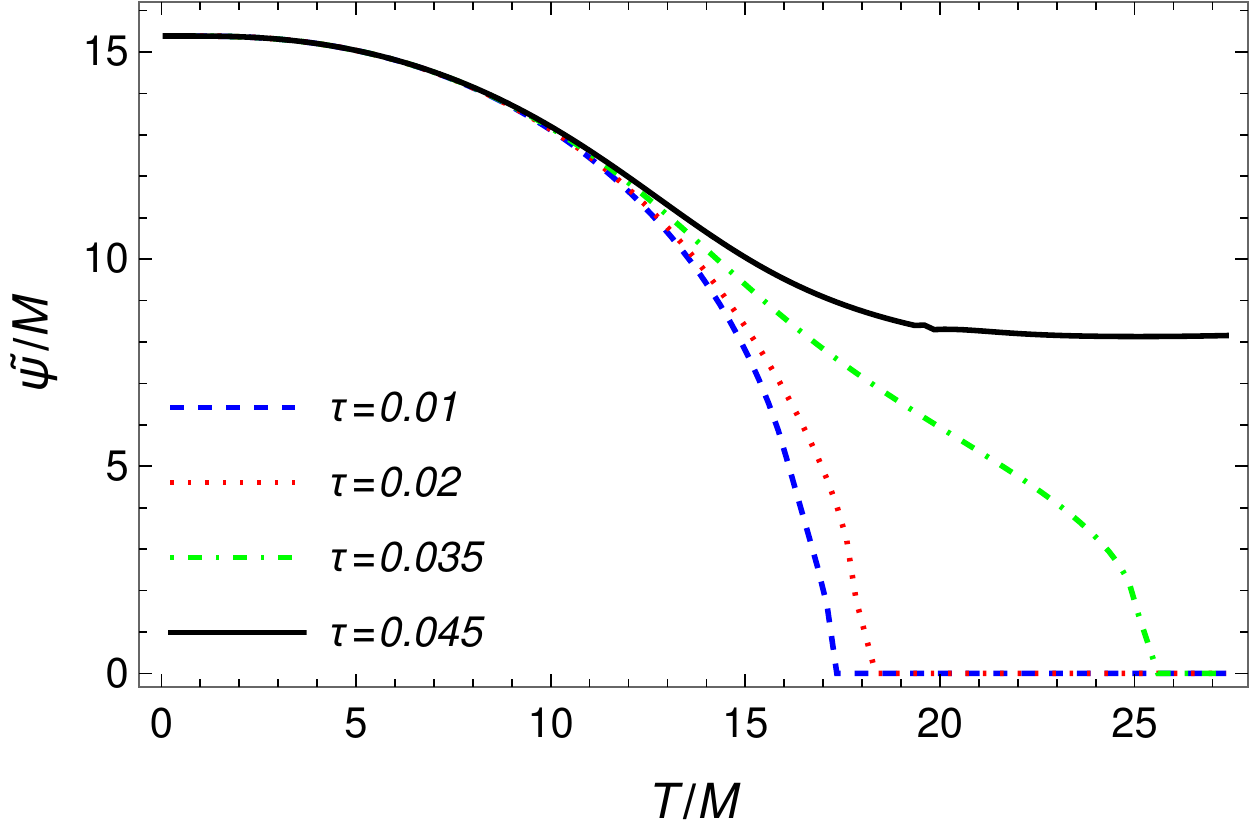}}
\caption{The behavior of the VEV of the fields, $\tilde{\phi}$ (panel
  a) and $\tilde{\psi}$ (panel b), as a function of temperature and
  for some representative values for the constant parameter $\tau_i$ in the charge density 
expression Eq.~(\ref{ratio-ns}). The parameters are such that $m_\phi^2 = m_\psi^2 = - m^2 <
  0$, $M=m$, $\tau_\phi = \tau_\psi=\tau$, $\lambda_\phi=0.018$,
  $\lambda_\psi=0.6$, and $\lambda=-0.03$.}
\label{fig10}
\end{figure}
\end{center}

In {}Fig.~\ref{fig10} we now consider the case where both fields are
initially in their broken symmetry states, i.e., $\tilde{\phi}\neq 0$
and $\tilde{\psi}\neq 0$ at $T=0$ and $\mu_\phi=\mu_\psi=0$. {}For the
parameters considered, in the absence of finite charge densities, the
transition pattern expected is the one shown in {}Fig.~\ref{fig7},
i.e., SNR in the direction of $\phi$ and usual SR transition in the
direction of $\psi$. {}From {}Fig.~\ref{fig10} we now see that besides
the $\phi$ field remaining in a SNR state,  now the finite charge
density also induces a SNR in the direction of the $\psi$ field. {}For
the parameters considered in {}Fig.~\ref{fig10}, for a charge density
over entropy density ratio $\tau \gtrsim 0.037$, we find that it
ceases to exist a critical temperature for SR in the direction of
$\psi$ and the field remains in a SNR state.

As a note to be remarked here, even though we have in this part of the
work continued to work with a negative value for the intercoupling
$\lambda$, in the presence of finite charge
densities both situations shown in {}Figs.~\ref{fig9} and \ref{fig10}
are still realized. ISB and SNR happens independently of the sign of
the coupling between $\phi$ and $\psi$. This happens exclusively
because of the effect of considering large enough densities.  Our
results then generalize to the two-field case the situation found in
Ref.~\cite{Benson:1991nj}, where it was studied for the one-field
case.

\section{Application to condensations of kaons}
\label{kaons}

As one of the possible applications of the methods and results studied
in this paper, one can consider the problem of the condensation of
kaons in QCD. Let us start by briefly reviewing the role of kaons in
the so-called CFL phase of QCD at high densities and how it can be
modeled with a system analogous to the model we have studied in the
previous sections. 

In QCD with three degenerate flavors and at asymptotically high
density and low temperatures, the ground state  for the quark matter
is supposed to be in the so-called CFL
state~\cite{Alford:1998mk,Alford:2007xm}, with diquark
condensates. This is a color superconducting state, where the quarks
can pair and form Cooper pairs, similar to what happens to electrons
in a condensed matter superconductor material. In this situation, that
can happen at sufficiently large densities, the original symmetry
group $SU(3)_c \times SU(3)_L \times SU(3)_R \times U(1)_B$, with the
color gauge group $SU(3)_c$ and the chiral symmetry group $SU(3)_L
\times SU(3)_R $, is broken down to $SU(3)_{c+L+R}$. The residual
group locks the rotations of color with rotations of the flavor, since
$SU(3)_{c+L+R}$ is a linear combination of the generators of the
original group.  The breakdown of the chiral symmetry gives origin to
an octet of pseudo-Goldstone modes and a singlet mode, which comes
from the breakdown of the baryon-number group  $U(1)_B$. The latter
group is called the {\it superfluid mode}, since it is responsible for
the superfluidity of the CFL phase.  At high densities all of the
modes can be regarded as approximately massless by neglecting the
masses of the quarks (e.g., the quarks up, down and strange).  This is
a rather different situation than that one at moderated densities,
when the chiral symmetry is explicitly broken and the superfluid mode
is the only one that remains massless, while the meson modes acquire
mass.  This scenario, of moderate densities (about the order of
several times the nuclear density), is the one expected for example in
the interior of neutron stars~\cite{Lee:1996ef} and the order of the
expected value for the chemical potential is about $500$ MeV. The
lightest mesons, except the massless superfluid mode, are the charged
and neutral kaons $K^+$, $K^-$ and $K^0$, $\bar{K}^0$, while the
strange quark, for example, has its mass in somewhere between the
current quark mass, $100$ MeV and the constituent quark mass,  $500$
MeV.  Likewise, it is to be noted that for a boson chemical potential
larger than its vacuum mass, the boson will suffer a Bose
condensation. This is what is supposed to happen with kaons in a dense
medium. The low in-medium kaon mass can, thus, lead to the formation
of a kaon condensate. 

As the symmetry-breaking pattern explained above is the same as in
vacuum QCD, it is expected the low-energy properties of the CFL phase
to be well described in terms of an effective chiral Lagrangian at
high densities for the octet of (pseudo-) Goldstone modes and the
superfluid mode.  The condensation of kaons can then be described by
an effective chiral Lagrangian density for the mesons in the CFL phase
as given by~\cite{Alford:2007qa,Andersen:2008tn}
\begin{eqnarray} 
\mathcal{L} &=& \frac{f_\pi^2}{4} {\rm tr} \left[\left(\partial_0
  \Sigma + [A,\Sigma]\right) \left(\partial_0 \Sigma^\dagger -
         [A,\Sigma]^\dagger\right) \right.  \nonumber \\ &-&
         \left. v_\pi^2 \left(\partial_i \Sigma\right)\left(\partial_i
         \Sigma^\dagger\right)\right]  \nonumber \\ &+& \frac{a
  f_\pi^2}{2} \det M\, {\rm
  tr}\left[M^{-1}(\Sigma+\Sigma^\dagger)\right] +\ldots, \nonumber \\
\label{chiralL}
\end{eqnarray}
where $\Sigma$ denotes the meson field, $\lambda^a$ are the Gell-Mann
matrices, $f_\pi$, $v_\pi$, and $a$ are constants, which can be found
by appropriate
matchings~\cite{Son:1999cm,Kaplan:2001qk,Schafer:2002ty}. In
Eq.~(\ref{chiralL}) $\mu_Q$ is the chemical potential for electric
charge and $\phi^a$ describes the octet of Goldstone bosons.  The
ellipses in Eq.~(\ref{chiralL}) stand for higher order terms in the
chiral Lagrangian density expansion. The matrix $A$ acts as a zeroth
component of a gauge field, which can be expressed in terms of the
diagonal matrices $Q=diag(2/3,-1/3,-1/3)$ and $M=diag(m_u,m_d,m_s)$,
with the chemical potential $\mu_Q$ and the baryon chemical potential
$\mu$ related by $A = \mu_Q Q - M^2/(2 \mu)$.

Using perturbative calculations for QCD at high densities, it is
possible to determine the parameters $f_\pi$, $v_\pi$ and $a$.  By
writing the meson field $\Sigma$ as
\begin{eqnarray}
 \Sigma = e^{i  \lambda^a \phi^a/f_\pi},
\end{eqnarray}
and expanding to fourth order in the meson fields, from the Lagrangian
density (\ref{chiralL}), the effective Euclidean Lagrangian density
for the kaons can be written in the form~\cite{Alford:2007qa}
\begin{eqnarray}
 \mathcal{L} &=& \left[\left(\frac{\partial}{\partial \tau} +
   \mu_1\right)\Phi_1^*\right] \left[\left(\frac{\partial}{\partial
     \tau} - \mu_1\right)\Phi_1\right] + (\partial_i
 \Phi_1^*)(\partial_i \Phi_1)  \nonumber \\ &+&
 \left[\left(\frac{\partial}{\partial \tau} +
   \mu_2\right)\Phi_2^*\right]\left[\left(\frac{\partial}{\partial
     \tau}- \mu_2\right)\Phi_2\right] + (\partial_i
 \Phi_2^*)(\partial_i \Phi_2) \nonumber\\ &+& m_1^2 \Phi_1^* \Phi_1 +
 m_{2}^2 \Phi_2^* \Phi_2 + \beta_1 (\Phi_1^* \Phi_1)^2 +\beta_2
 (\Phi_2^* \Phi_2)^2  \nonumber \\ &+& 2\alpha (\Phi_1^* \Phi_1)
 (\Phi_2^* \Phi_2),
\label{chiralphi1phi2}
\end{eqnarray}
where the complex doublet scalar field $(\Phi_1,\Phi_2)$ can be
identified with the charged and the neutral kaons,
$(K_+,K_0)=(\Phi_1,\Phi_2)$, with the chemical potentials $\mu_1$ and
$\mu_2$ associated with the conserved charges for  $\Phi_1$ and
$\Phi_2$, respectively. The effective model for the kaons, when
expressed in the form of Eq.~(\ref{chiralphi1phi2}), is just of the
same form as the model we have studied here, in terms of the Euclidean
action Eq.~(\ref{SEucl}), and  by identifying, e.g., $(\phi,\psi)$
with $(\Phi_1,\Phi_2)$, or, equivalently, with $(K_+,K_0)$, with also
$m_\phi\equiv m_1$, $m_\psi \equiv m_2$, $\beta_1 \equiv
\lambda_\phi/6$, $\beta_2 \equiv \lambda_\psi/6$ and $\alpha\equiv
\lambda/2$.

In Refs.~\cite{Alford:2007qa,Andersen:2008tn,Tran:2008mvg} the kaon
condensation problem with the effective model
Eq.~(\ref{chiralphi1phi2}) was studied using the CJT method. Here, we
want to compare the same results but in terms of the OPT. This also
gives us an opportunity for contrasting our results with a different
nonperturbative method. To facilitate this comparison, we will make
use of the same parameters considered for example in
Ref.~\cite{Tran:2008mvg}. It should be noted that in
Ref.~\cite{Alford:2007qa}, the authors, for simplicity, disregarded
the quantum contributions in their analysis and kept only the thermal
integrals in the expressions. The authors in
Ref.~\cite{Andersen:2008tn} made use of slight different choice of
parameters than the ones used in
Refs.~\cite{Alford:2007qa,Tran:2008mvg}, but in all the cases the
results obtained were qualitatively  similar. Since in
Ref.~\cite{Tran:2008mvg} the authors kept all the correction terms
(quantum and thermal) in their expressions, we find easily to compare
their results with ours. In the parameters considered in
Ref.~\cite{Tran:2008mvg}, we have for instance that $m_1= 5$ MeV,
$m_2=4$ MeV, $\mu_1=\mu_2=4.5$ MeV, $\beta_1=0.0048$, $\beta_2=0.005$
and $\alpha=0.046$. The authors of Ref.~\cite{Tran:2008mvg} have also
worked with a fixed value for the renormalization scale $M$ as $M=4.5$
MeV.  Given that $m_2<\mu_2$, condensation of $K_0$, which is
associated with the $\Phi_2$ complex scalar field in
Eq.~(\ref{chiralphi1phi2}), is expected to happen. 

\begin{center}
\begin{figure}[!htb]
\includegraphics[width=7.5cm]{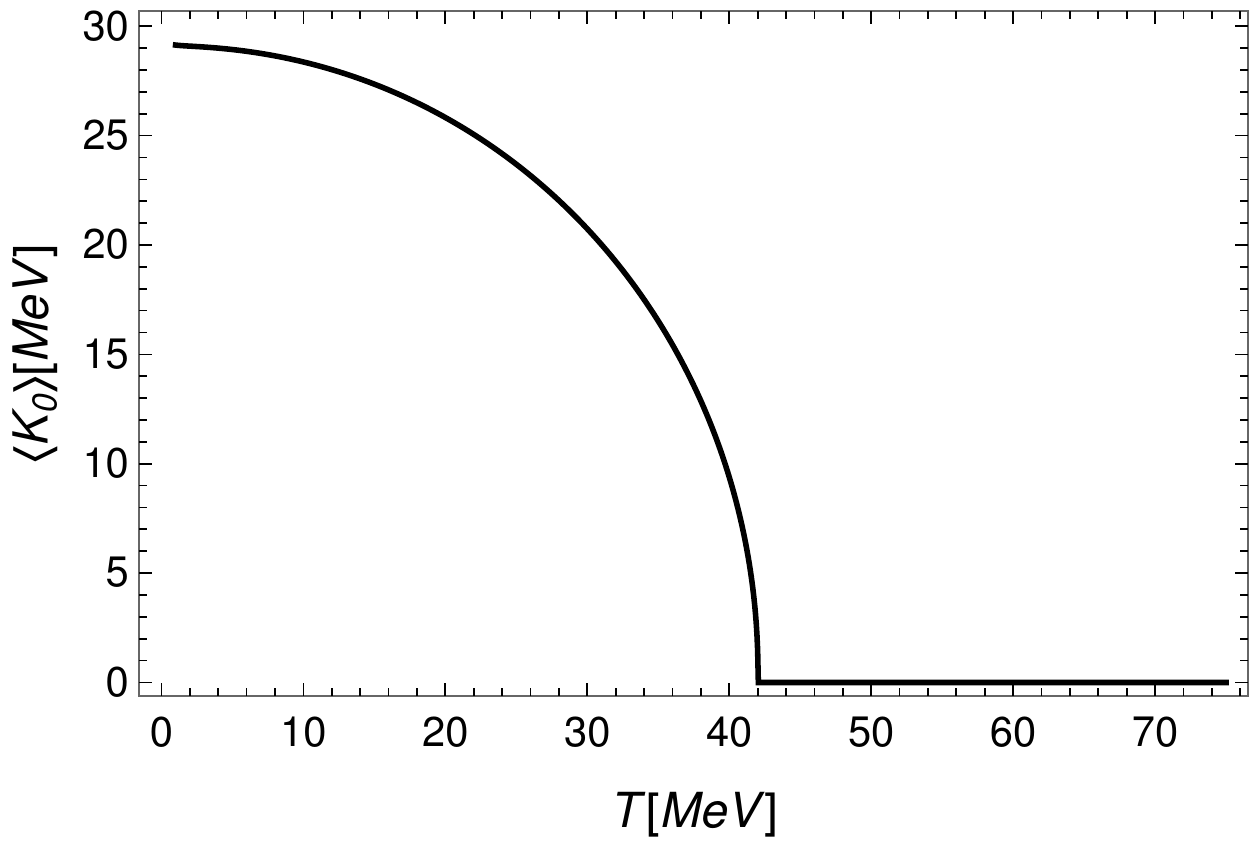}
\caption{The neutral kaon $K_0$ ($\Phi_2$) VEV as a function of the
  temperature for the parameters given in the text.}
\label{fig11}
\end{figure}
\end{center}

\begin{center}
\begin{figure}[!htb]
\includegraphics[width=7.5cm]{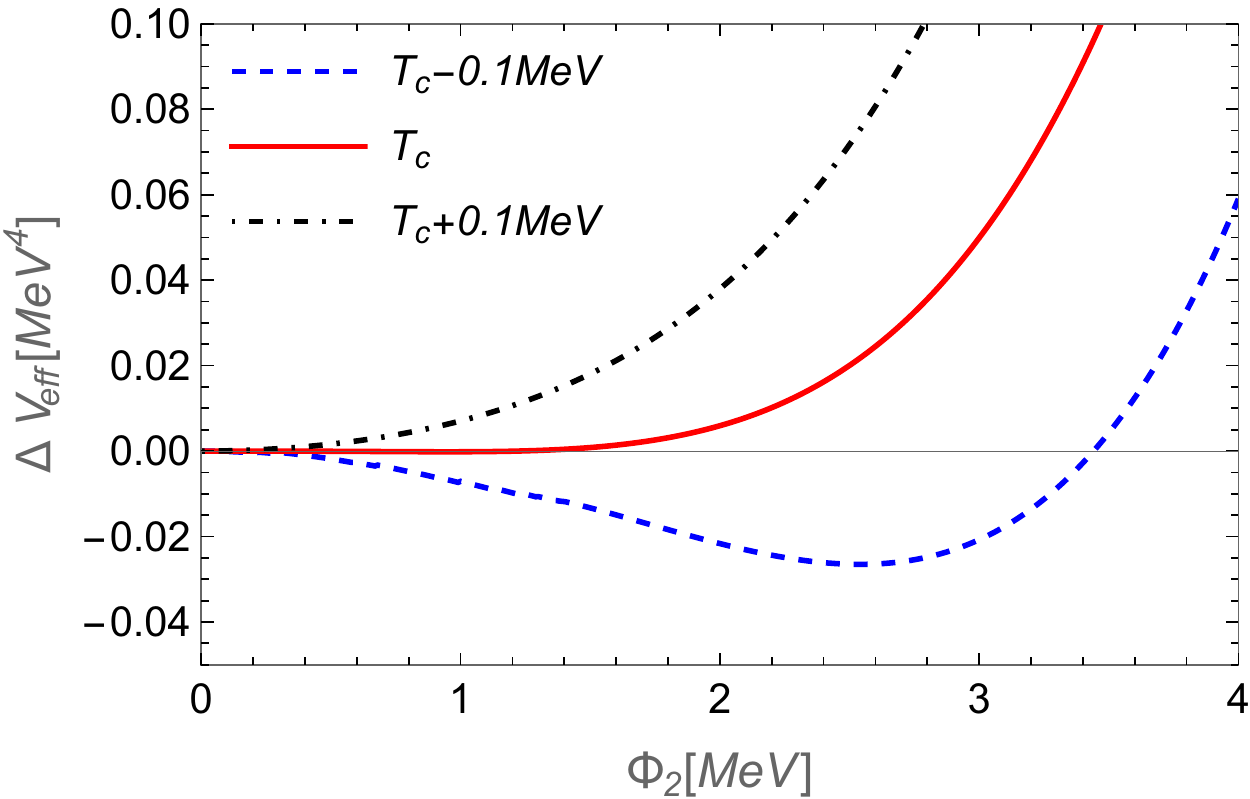}
\caption{The thermodynamic potential (subtracted by its value in the
  vacuum) as a function of the VEV in the $\Phi_2$ field direction.}
\label{fig12}
\end{figure}
\end{center}

In {}Fig.~\ref{fig11} we show how the VEV associated with the $\Phi_2$
field (e.g., with $K_0$) changes with the temperature. The transition
temperature found in the context of the the OPT is $T_{c,K_0}\simeq
42$ MeV. This result completely agrees with the one found in figure 2
of Ref.~\cite{Tran:2008mvg}. {}For comparison purposes, we also show
in {}Fig.~\ref{fig12} the way that the effective thermodynamic
potential changes along the $\Phi_2$ field direction.  We have
considered a variation of $0.1$ MeV around the critical
temperature. The phase transition is found to be of second-order,
which is also in agreement to the results shown in {}Fig.~3 of
Ref.~\cite{Tran:2008mvg}.

\begin{center}
\begin{figure}[!htb]
\includegraphics[width=7.5cm]{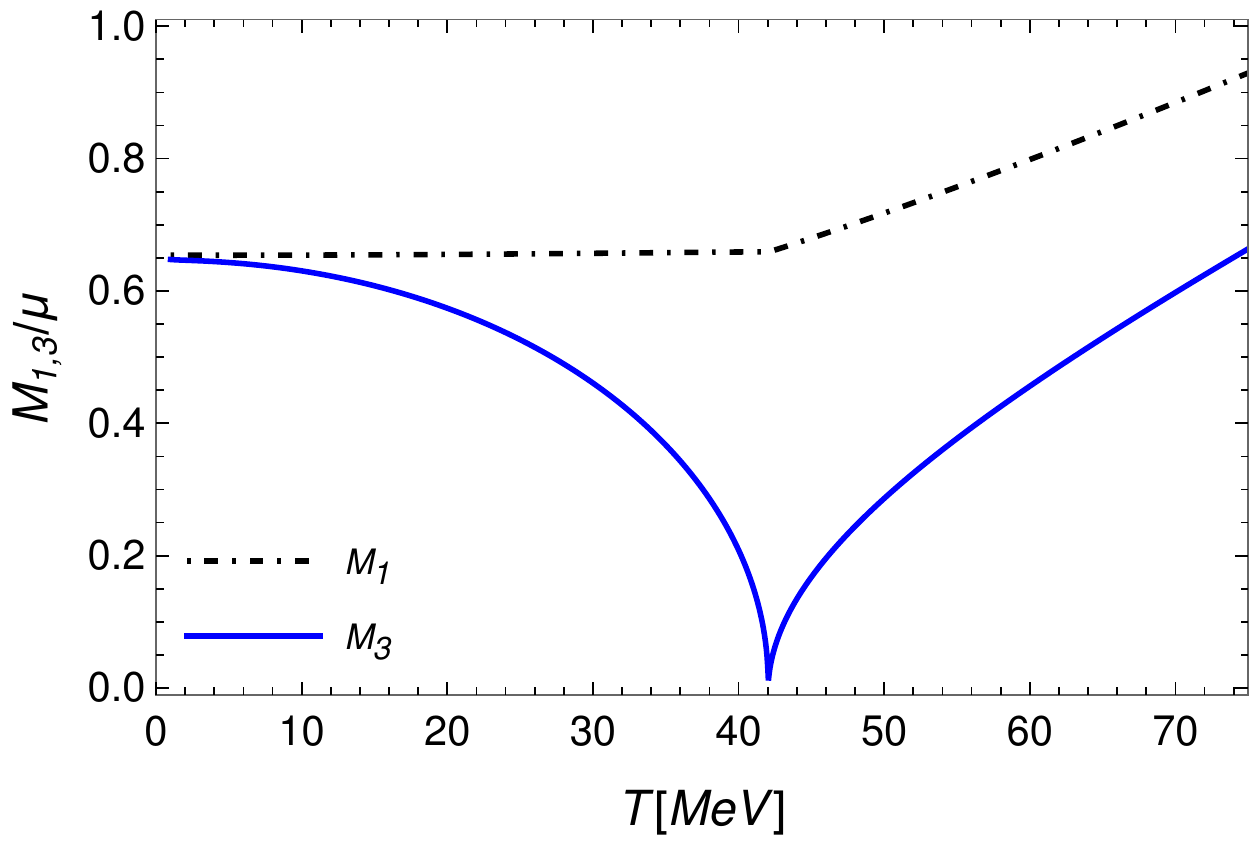}
\caption{The self-consistent kaon masses $M_1$ and $M_3$ in the
  notation of Refs.~\cite{Alford:2007qa,Tran:2008mvg}.  Here, the
  masses are normalized by the value $\mu=4.5$ MeV.}
\label{fig13}
\end{figure}
\end{center}

The self-consistent kaon masses discussed in
Refs.~\cite{Alford:2007qa,Andersen:2008tn,Tran:2008mvg} are
equivalently expressed in terms of the Eqs.~(\ref{mphi2eff}) and
(\ref{mpsi1eff}), which we have defined previously. In the notation
used in  Refs.~\cite{Alford:2007qa,Andersen:2008tn,Tran:2008mvg}, they
can be identified with their  self-consistent masses $M_1$ and $M_3$
(note that in Ref.~\cite{Andersen:2008tn}, they identify $K_0$ with
$\Phi_1$ and $K_+$ with $\Phi_2$ instead).  These self-consistent
masses are shown in {}Fig.~\ref{fig13}.

An issue of particular importance discussed in
Refs.~\cite{Alford:2007qa,Andersen:2008tn,Tran:2008mvg}, was the
difficulty of the Goldstone theorem to be satisfied in the CJT
formalism. Slight variations of implementation of the CJT in those
references have been proposed in order for the Goldstone theorem to be
satisfied at least in an approximate form. As we have already
discussed in the previous sections, in the OPT formalism, as seen
explicitly at the order of the OPT considered in the previous section,
the Goldstone theorem is satisfied exactly.  As we have already
demonstrated the validity of the Goldstone theorem through the mass
eigenvalues defined in the previous sections and exemplified, e.g., by
{}Figs.~\ref{fig6} and \ref{fig8}, we refrain here from doing the same
thing, but using simply a different set of parameters.

{}Finally, we can also implement the constrain of charge neutrality
for the kaon system in the same manner as discussed in
Refs.~\cite{Alford:2007qa,Andersen:2008tn,Tran:2008mvg}. This is of
particular importance when applying the results for bulk matter in
compact stars, where there is overall color and electrical charges
neutrality.  By assuming an ideal Fermi gas for the background of
electrons, charge neutrality can be implemented by adding to the
Lagrangian density Eq.~(\ref{chiralphi1phi2}) the contribution from
the electrons,
\begin{equation}
{\cal L}_{\rm electrons} = \bar\psi_e(\not\!\partial + \gamma^0 \mu_Q
Q_e - m_e)\psi_e,
\label{elecrons}
\end{equation}
where $\psi_e$ here denotes the Dirac field for the electron, $Q_e$
the electron charge, $\mu_Q$ the chemical potential and $m_e$ the
electron mass. The additional contribution for the thermodynamic
potential, in the high temperature approximation $T \gg m_e$, is
simply
\begin{equation}
V_{\rm eff, electrons} \simeq - \frac{\mu_Q^4}{12\pi^2}
-\frac{\mu_Q^2}{6}T^2 - \frac{7 \pi^2}{180}T^4,
\end{equation}
for which the electron charge density gives,
\begin{eqnarray}
n_e = \frac{\mu_Q^3}{3 \pi^2} + \frac{\mu_Q}{3}T^2.
\end{eqnarray}
The overall charge neutrality for the system imposes the constraint
(recalling the we are associating the complex scalar field $\Phi_1$
with the charged kaon)
\begin{eqnarray}
n_1 + n_e = 0.
\end{eqnarray}

\begin{center}
\begin{figure}[!htb]
\includegraphics[width=7.5cm]{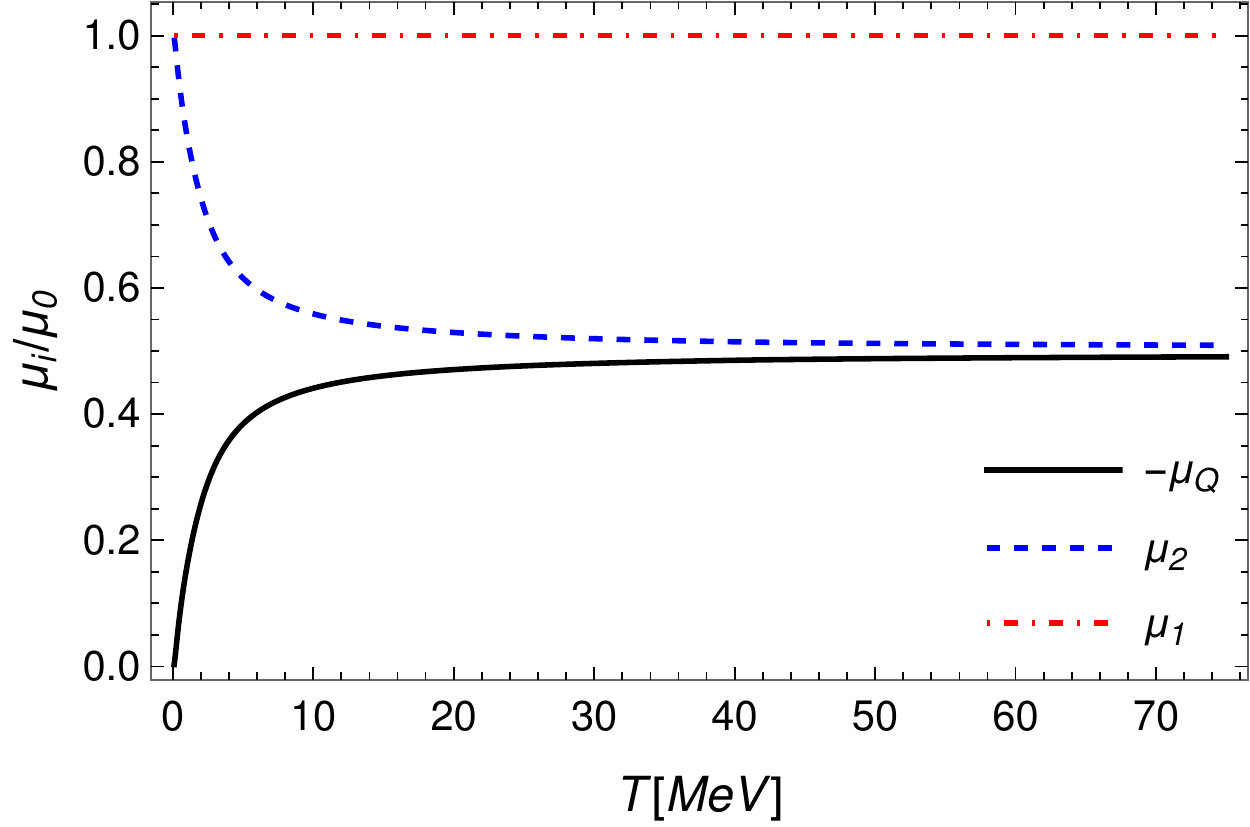}
\caption{The chemical potentials for the kaons and electron when
  imposing charge neutrality.}
\label{fig14}
\end{figure}
\end{center}

In {}Fig.~\ref{fig14}, we show the chemical potentials associated with
the kaons and electron as a  function of the temperature. This result
can be contrasted for instance with the {}Fig.~7 in
Ref.~\cite{Alford:2007qa} or {}Fig.~1 in Ref.~\cite{Tran:2008mvg}.

\section{Conclusions}
\label{conclusions}

In this paper we have studied the question of achieving symmetry
inversion and symmetry persistence, ISB and SNR, respectively, at high
temperatures when finite charges are taken into account.  The results
were studied by making use of the nonperturbative OPT method, which
has already been used successfully before in many other different
contexts. 

We have shown that the chemical potential for the fields tends to
favor both ISB and SNR phenomena in the context of a two complex
scalar field system. This happens such that, for instance, the
critical temperature for ISB becomes smaller the larger are the
chemical potentials for the fields. When studying the same system at
finite density charges, we have followed the pioneering work
considered in Refs.~\cite{Bernstein:1990kf,Benson:1991nj}, which
studied the one complex scalar field case. By working with a fixed
charge density over entropy density ratio, which is motivated for
example in cosmological settings, we have demonstrated that charges
densities further facilitate both ISB and SNR. The finite density
effects allow for both fields to experience ISB or SNR, which is not
allowed in the absence of conserved charges. Our results, thus, extend
to multiple coupled complex scalar field systems the study originally
performed only in the one complex field case.

{}Finally, as an application of our results and the OPT method used in
our study, we have considered the condensation of kaons, as expected,
e.g., in a CFL phase of QCD at large densities.  We have contrasted
our results with similar ones previously considered in the literature,
but which  made use of the CJT nonperturbative method. Our results
with the OPT, besides of comparing favorable with those obtained with
the CJT method, have the additional advantage of being simpler in
implementing and at the same time fully satisfying the Goldstone
theorem, which has been a particular issue in the other
nonperturbative methods.

\begin{acknowledgments}

R.O.R. would like to acknowledge the McGill University Physics Department 
for the hospitality. 
The authors acknowledge financial  support from  Coordena\c{c}\~ao de
Aperfei\c{c}oamento de Pessoal de N\'{\i}vel Superior (CAPES) -
Finance Code 001 and by research grants from Conselho Nacional de
Desenvolvimento Cient\'{\i}fico e Tecnol\'ogico (CNPq), Grant
No. 307286/2021-5 (R.O.R) and No. 309598/2020-6 (R.L.S.F.),  from
Funda\c{c}\~ao Carlos Chagas Filho de Amparo \`a Pesquisa do Estado do
Rio de Janeiro (FAPERJ), Grant No. E-26/201.150/2021 (R.O.R.) and from
FAPERGS Grants Nos. 19/2551-0000690-0 and 19/2551-0001948-3
(R.L.S.F.).

\end{acknowledgments}

\appendix

\section{The ETP to first order in the OPT}
\label{appA}

Let us explicitly derive each one of the terms in {}Fig.~\ref{fig1}
and which contribute to the ETP at first order in the OPT.  We work in
Euclidean space, with momentum square denoted by $P^2= p_4^2 + {\bf
  p}^2$, where $p_4\equiv \omega_n= 2\pi n T$, $n\in \mathbb{Z}$ and
$\omega_n$ are the Matsubara's frequencies for bosons.  All momenta
integrals are evaluated in dimensional regularization in the
$\overline{\mathrm{MS}}$ scheme. Hence, the momentum integrals in the
loop contributions at finite temperature and chemical potential are
represented by
\begin{equation}
\sum_{P}\!\!\!\!\!\!\!\!\int \equiv
T\sum_{p_{4}=\omega_n+i\mu_i}\left( \frac{e^{\gamma_E }M^{2}}{4\pi }
\right)^{\epsilon }\int \frac{d^{d}p}{\left( 2\pi \right)^{d}},
\label{sumT}
\end{equation}
where $\mu_i\equiv \mu_{\phi(\psi)}$ is the chemical potential
associated with the $\phi(\psi)$ field.  The  divergent vacuum
momentum integral terms are regularized in the
$\overline{\mathrm{MS}}$ scheme, with $d=3-2\epsilon $, $\gamma_{E}$
is the Euler-Mascheroni constant, $\gamma_E \simeq 0.577$, and $M$ is
the arbitrary mass  regularization scale. The sum in Eq.~(\ref{sumT})
is performed over the Matsubara's frequencies. 

Performing the sum over the Matsubara's frequencies and the momentum
integrals in dimensional regularization, we have for instance that
\begin{equation}
\sum_{P}\!\!\!\!\!\!\!\!\int \ln \left(P^{2}+\Omega^{2}\right) =
-\frac{\Omega^4}{2(4\pi)^2}\frac{1}{ \epsilon} + Y(\Omega,T,\mu),
\label{intlog}
\end{equation}
and
\begin{equation}
\sum_{P}\!\!\!\!\!\!\!\!\int \frac{1}{P^{2}+\Omega^{2}} =
-\frac{\Omega^2}{(4\pi)^2}\frac{1}{ \epsilon} + X(\Omega,T,\mu),
\label{intG}
\end{equation}
where
\begin{equation}
Y(\Omega,T,\mu) = \frac{\Omega^4}{2(4\pi)^2}\left[  \ln\left(
  \frac{\Omega^2}{M^2}\right) -\frac{3}{2}\right] +J_B(\Omega,T,\mu),
\label{YTmu}
\end{equation}
and 
\begin{equation}
X(\Omega,T,\mu) = \frac{\Omega^2}{(4\pi)^2}\left[  \ln\left(
  \frac{\Omega^2}{M^2}\right) -1\right] +I_B(\Omega,T,\mu),
\label{XTmu}
\end{equation}
with $J_B(\Omega,T,\mu)$ and $I_B(\Omega,T,\mu)$ denoting the thermal
integrals, are defined as
\begin{eqnarray}
J_B(\Omega ,T,\mu ) &=&   \frac{T^4}{2\pi^{2}}\int_{0}^{\infty
}dzz^{2} \left\{\ln \left[ 1-e^{-\sqrt{z^{2}+
      \frac{\Omega^{2}}{T^{2}} } + \frac{\mu}{T}} \right]  \right.
\nonumber \\  &+&   \left.  \ln \left[ 1-e^{-\sqrt{z^{2}+
      \frac{\Omega^{2}}{T^{2}} } - \frac{\mu}{T}} \right]
\right\}, \label{JB}
\end{eqnarray}
and
\begin{eqnarray}
I_B(\Omega ,T,\mu ) &\equiv& \frac{\partial J_B(\Omega ,T,\mu
  )}{\partial \Omega^ 2}  \nonumber \\ &=& \frac{T^{2}}{4\pi^2
}\int_{0}^{\infty }dz \frac{z^{2}}{\sqrt{z^{2}+
    \frac{\Omega^{2}}{T^{2}}}} \nonumber\\ &\times& \left[\frac{1}{
    e^{ \sqrt{ z^{2}+  \frac{\Omega^2}{T^2} } +\frac{\mu}{T} } -1    }
  + \frac{1}{    e^{ \sqrt{ z^{2}+  \frac{\Omega^{2}}{T^{2}} }
      -\frac{\mu}{T} } -1    } \right].  \nonumber \\
\label{IB}
\end{eqnarray}
Note that in the notation of Harber and Welson,
Ref.~\cite{Haber:1981tr}, the thermal integrals in Eqs.~(\ref{JB}) and
(\ref{IB}), are given, respectively, by
\begin{eqnarray}
J_B(\Omega,T,\mu) & \equiv &  -\frac{\Gamma(5)}{3}\frac{T^4}{\pi^{2}}
h_5^e(y,r)
\label{JBh}
\end{eqnarray}
and
\begin{eqnarray}
I_B(\Omega,T,\mu) & \equiv &  \Gamma(3)\frac{T^2}{2\pi^{2}} h_3^e(y,r)
\label{IBh}
\end{eqnarray}
where $y \equiv \Omega/T$ and $r \equiv \mu/\Omega$. As shown in
Ref.~\cite{Haber:1981tr}, the thermal functions $h_{2n+1}^e(y,r)$ have
a high temperature expansion given by
\begin{eqnarray}
 h^e_{2n+1}(y,r) &=& \frac{\pi y^{2n-1}}{2\Gamma(2n+1)}(-1)^n
 (1-r^2)^{n-\frac{1}{2}}
 \nonumber\\ &+&\frac{(-1)^n}{2[\Gamma(n+1)]^2}
 \left(\frac{y}{2}\right)^{2n} \left[ \ln \left(\frac{y}{4\pi}\right)
   + \frac{\gamma_E}{2}  \right.  \nonumber \\ &-& \left.
   \frac{\psi(n+1)}{2} + nr^2\; {}_3F_2(1,1,1-n;\frac{3}{2},2;r^2)
   \right]
 \nonumber\\ &+&\frac{(-1)^n}{2\Gamma(n+1)}\left(\frac{y}{2}\right)^{2n}
 \sum^{\infty}_{k=1}(-1)^k \left(\frac{y}{4 \pi }\right)^{2k}
 \nonumber \\ &\times&  \frac{\Gamma(2k+1) \, \zeta(2k+1)}
           {\Gamma(k+1) \Gamma(k+1+n)}\;
           {}_2F_1(-k,-n-k;\frac{1}{2};r^2)
           \nonumber\\ &+&\frac{1}{2\Gamma(n+1)}
           \sum^{n-1}_{k=0}(-1)^k \left(\frac{y}{2}\right)^{2k}
           \ \nonumber \\ &\times& \frac{\Gamma(n-k) \,
             \zeta(2n-2k)}{\Gamma(k+1)} \;
                       {}_2F_1(-k,n-k;\frac{1}{2};r^2)  , \nonumber \\
\end{eqnarray}
where $\psi(n)$ is the Digama function,  $_2F_1(a,b;c;z)$ and
$_3F_2(a,b,c;d,e;z)$ are hypergeometric functions, $\zeta(n)$ is the
Riemann zeta function and $\Gamma(n)$ is the Gamma function.

Individually, each diagram shown in {}Fig.~\ref{fig1}, after
performing the sum over the Matsubara's frequencies  and the momentum
integrals for the vacuum terms, is then explicitly given by
\begin{eqnarray}
V_{\rm eff}^{(a)}&=& \sum_{P}\!\!\!\!\!\!\!\!\int \ln
\left(P^{2}+\Omega_{\phi }^{2}\right) \nonumber\\ &=&
-\frac{\Omega_{\phi }^{4}}{2\left( 4\pi \right)^{2}} \frac{1}{\epsilon
} + Y(\Omega_\phi,T,\mu_\phi), 
\label{Va}
\end{eqnarray}
\begin{eqnarray}
V_{\rm eff}^{(b)}&=& \sum_{P}\!\!\!\!\!\!\!\!\int \ln
\left(P^{2}+\Omega_{\psi }^{2}\right)
\nonumber\\ &=&-\frac{\Omega_{\psi }^{4}}{2\left( 4\pi \right)^{2}}
\frac{1}{\epsilon }+ Y(\Omega_\psi,T,\mu_\psi),
\label{Vb}
\end{eqnarray}
\begin{eqnarray}
V_{\rm eff}^{(c)}&=& -\delta \eta_{\phi
}^{2}\sum_{P}\!\!\!\!\!\!\!\!\int \frac{1}{P^{2}+\Omega_{\phi }^{2}}
\nonumber \\ &=&-\delta \eta_{\phi }^{2}\left[ -\frac{\Omega_{\phi
    }^{2}}{\left(4\pi \right)^{2}}\frac{1}{\epsilon
  }+X(\Omega_\phi,T,\mu_\phi) \right] ,
\label{Vc}
\end{eqnarray}
\begin{eqnarray}
V_{\rm eff}^{(d)}&=& -\delta \eta_{\psi
}^{2}\sum_{P}\!\!\!\!\!\!\!\!\int \frac{1}{P^{2}+\Omega_{\psi }^{2}}
\nonumber \\ &=&-\delta \eta_{\psi }^{2}\left[ -\frac{\Omega_{\psi
    }^{2}}{\left(4\pi \right)^{2}}\frac{1}{\epsilon
  }+X(\Omega_\psi,T,\mu_\psi)\right],
\label{Vd}
\end{eqnarray}
\begin{eqnarray}
V_{\rm eff}^{(e)}&=& \delta \frac{ \lambda_{\phi
}}{3}\phi_0^{2}\sum_{P}\!\!\!\!\!\!\!\!\int
\frac{1}{P^{2}+\Omega_{\phi }^{2}} \nonumber \\ &=&\frac{1}{3}\delta
\lambda_{\phi }\phi_0^{2}\left[- \frac{\Omega_{\phi }^{2}}{\left( 4\pi
    \right)^{2}}\frac{1}{\epsilon }+X(\Omega_\phi,T,\mu_\phi) \right]
,
\label{Ve}
\end{eqnarray}
\begin{eqnarray}
V_{\rm eff}^{(f)}&=& \delta
\frac{\lambda_\psi}{3}\psi_0^{2}\sum_{P}\!\!\!\!\!\!\!\!\int\frac{1}{P^{2}+\Omega_{\psi
  }^{2}}   \nonumber \\ &=& \frac{1}{3}\delta \lambda_{\psi
}\psi_0^{2}\left[- \frac{\Omega_{\psi }^{2}}{\left( 4\pi
    \right)^{2}}\frac{1}{\epsilon }+X(\Omega_\psi,T,\mu_\psi) \right]
,
\label{Vf}
\end{eqnarray}
\begin{eqnarray}
V_{\rm eff}^{(g)}&=& \delta\frac{\lambda}{2}
\psi_0^{2}\sum_{P}\!\!\!\!\!\!\!\!\int \frac{1}{P^{2}+\Omega_{\phi
  }^{2}} \nonumber \\ &=&\frac{1}{2}\delta \lambda \psi_0^{2}\left[-
  \frac{\Omega_{\phi}^{2}}{\left( 4\pi \right)^{2}}\frac{1}{\epsilon
  }+X(\Omega_\phi,T,\mu_\phi) \right] ,
\label{Vg}
\end{eqnarray}
\begin{eqnarray}
V_{\rm eff}^{(h)}&=& \delta
\frac{\lambda}{2}\phi_{0}^{2}\sum_{P}\!\!\!\!\!\!\!\!\int
\frac{1}{P^{2}+\Omega_{\psi }^{2}} \nonumber \\ &=&\frac{1}{2}\delta
\lambda \phi_0^{2}\left[- \frac{\Omega_{\psi}^{2}}{\left( 4\pi
    \right)^{2}}\frac{1}{\epsilon }+X(\Omega_\psi,T,\mu_\psi)  \right]
,
\label{Vh}
\end{eqnarray}
\begin{eqnarray}
V_{\rm eff}^{(i)}&=& \delta \frac{\lambda_\phi}{3}\left[
  \sum_{P}\!\!\!\!\!\!\!\!\int\frac{1}{P^{2}+\Omega_{\phi
    }^{2}}\right]^{2} \nonumber \\ &=&\delta \frac{\lambda_\phi}{3}
\left[ \frac{\Omega_\phi^4}{(4 \pi)^4}  \frac{1}{\epsilon^2}  -
  \frac{\Omega_\phi^2}{8\pi^2}  \frac{1}{\epsilon}
  X(\Omega_\phi,T,\mu_\phi) \right.  \nonumber \\ &+&
  \left. 2\frac{\Omega_\phi^4}{(4 \pi)^4} W(\Omega_\phi)   +
  X^2(\Omega_\phi,T,\mu_\phi)\right],
\label{Vi}
\end{eqnarray}
\begin{eqnarray}
V_{\rm eff}^{(j)}&=& \delta\frac{\lambda_\psi}{3}\left[
  \sum_{P}\!\!\!\!\!\!\!\!\int
  \frac{1}{P^{2}+\Omega_{\psi}^{2}}\right]^{2} \nonumber \\ &=&\delta
\frac{\lambda_\psi}{3} \left[ \frac{\Omega_\psi^4}{(4 \pi)^4}
  \frac{1}{\epsilon^2}  - \frac{\Omega_\psi^2}{8\pi^2}
  \frac{1}{\epsilon} X(\Omega_\psi,T,\mu_\psi) \right.  \nonumber
  \\ &+& \left. 2\frac{\Omega_\psi^4}{(4 \pi)^4} W(\Omega_\psi)   +
  X^2(\Omega_\psi,T,\mu_\psi)\right],
\label{Vj}
\end{eqnarray}
\begin{eqnarray}
V_{\rm eff}^{(k)} &=& \delta \lambda\left[
  \sum_{P}\!\!\!\!\!\!\!\!\int \frac{1}{P^{2}+\Omega_{\phi
    }^{2}}\right]  \left[ \sum_{P}\!\!\!\!\!\!\!\!\int
  \frac{1}{P^{2}+\Omega_{\psi }^{2}}\right]    \nonumber \\ &=&\delta
\lambda \left\{  \frac{\Omega_\phi^2\Omega_\psi^2}{(4 \pi)^4}
\frac{1}{\epsilon^2} \right.  \nonumber\\ &-& \left.
\frac{1}{(4\pi)^2\epsilon}\left[\Omega_\phi^2
  X(\Omega_\psi,T,\mu_\psi)+ \Omega_\psi^2 X(\Omega_\phi,T,\mu_\phi)
  \right] \right.  \nonumber \\ &+& \left. \frac{\Omega_\phi^2
  \Omega_\psi^2}{(4 \pi)^4} \left[W(\Omega_\phi)+W(\Omega_\psi)
  \right]   \right.  \nonumber\\ &+& \left. X(\Omega_\phi,T,\mu_\phi)
X(\Omega_\psi,T,\mu_\psi)   \right\},
\label{Vk}
\end{eqnarray}
where in the above expressions, we have also defined
\begin{equation}
W(\Omega) = \frac{1}{2}\left[ \ln \left(
  \frac{\Omega^{2}}{M^{2}}\right) -1\right]
^{2}+\frac{1}{2}+\frac{\pi^{2}}{12} .
\end{equation}

The divergent terms in Eqs.~(\ref{Ve})-(\ref{Vh}) can be eliminated by
introducing the mass counterterms in the  OPT Lagrangian density
Eq.~(\ref{OPTL}) by redefining the bare quadratic OPT masses as
$m_\phi^2 \to m_\phi^2 + \Delta m_\phi^2$ and $m_\psi^2 \to m_\psi^2 +
\Delta m_\psi^2$, where the counterterms $\Delta m_\phi^2$ and
$\Delta m_\psi^2$ are given, respectively, by
\begin{eqnarray}
\Delta m_\phi^2 &=& \frac{1}{16 \pi^2 \epsilon}\left( \frac{2 \delta
  \lambda_\phi}{3}\Omega_\phi^2 +  \delta \lambda \Omega_\psi^2
\right), 
\label{Dmphi}
\\ \Delta m_\psi^2 &=&\frac{1}{16 \pi^2 \epsilon}\left( \frac{2 \delta
  \lambda_\psi}{3}\Omega_\psi^2 +  \delta \lambda \Omega_\phi^2
\right). \label{Dmpsi}
\end{eqnarray}
These mass counterterms then give the explicit additional
contributions at order $\delta$,
\begin{eqnarray}
V_{\rm eff}^{(l)}&=& \frac{\Delta m_\phi^2}{2}  \phi_0^2,
\label{Vl}
\\ V_{\rm eff}^{(m)}&=& \frac{\Delta m_\psi^2}{2}  \psi_0^2.
\label{Vm}
\end{eqnarray}

The mass counterterms also enter as additional vertices, just like in
the standard perturbation theory case, which then lead to the
additional loop contributions at order $\delta$,
\begin{eqnarray}
V_{\rm eff}^{(n)}& =& \frac{1}{16 \pi^2 \epsilon}\left( \frac{2 \delta
  \lambda_\phi}{3}\Omega_\phi^2 +  \delta \lambda \Omega_\psi^2
\right)\sum_{P}\!\!\!\!\!\!\!\!\int \frac{1}{P^{2}+\Omega_{\phi }^{2}}
\nonumber \\ &=& \left( \frac{2 \delta \lambda_\phi}{3}\Omega_\phi^2 +
\delta \lambda \Omega_\psi^2  \right)\left[  -\frac{\Omega_\phi^2}{(4
    \pi)^4}  \frac{1}{\epsilon^2}  \right.  \nonumber\\ &+&
  \left. \frac{1}{(4\pi)^2 \epsilon}   X(\Omega_\phi,T,\mu_\phi)
  -\frac{\Omega_\phi^2}{(4\pi)^4} W(\Omega_\phi)\right], 
\label{Vn}
\end{eqnarray}
\begin{eqnarray}
V_{\rm eff}^{(o)}& =& \frac{1}{16 \pi^2 \epsilon}\left( \frac{2 \delta
  \lambda_\psi}{3}\Omega_\psi^2 +  \delta \lambda \Omega_\phi^2
\right)\sum_{P}\!\!\!\!\!\!\!\!\int \frac{1}{P^{2}+\Omega_{\psi }^{2}}
\nonumber \\ &=&\left( \frac{2 \delta \lambda_\psi}{3}\Omega_\psi^2 +
\delta \lambda \Omega_\phi^2  \right)\left[  -\frac{\Omega_\psi^2}{(4
    \pi)^4}  \frac{1}{\epsilon^2}  \right.  \nonumber\\ &+&
  \left. \frac{1}{(4\pi)^2 \epsilon}   X(\Omega_\psi,T,\mu_\psi)
  -\frac{\Omega_\psi^2}{(4\pi)^4} W(\Omega_\psi)\right] .
\label{Vo}
\end{eqnarray}
{}Finally, the remaining divergences are all vacuum terms, which can
be canceled by adding to the OPT Lagrangian density Eq.~(\ref{OPTL})
a vacuum renormalization counterterm,
\begin{eqnarray}
V_{\rm eff}^{(p)}& \equiv& \Delta V = \frac{\Omega_{\phi }^{4} +
  \Omega_{\psi }^{4}}{2\left( 4\pi \right)^{2}} \frac{1}{\epsilon }
\nonumber \\ &-& \frac{\delta}{\left(4\pi \right)^{2} \epsilon }\left[
  \eta_\phi^2 \Omega_{\phi }^{2}+  \eta_\psi^2 \Omega_{\psi }^{2}
  \right] \nonumber \\ &+& \frac{\delta}{(4 \pi)^4\epsilon^2}
\left[\frac{\lambda_\phi}{3} \Omega_\phi^4+ \frac{\lambda_\psi}{3}
  \Omega_\psi^4  + \lambda \Omega_\phi^2\Omega_\psi^2\right].
\label{Vp}
\end{eqnarray} 
Note that at first order in the OPT there are no vertex counterterms
that are required. Vertex counterterms do appear though  when carrying
out the OPT at second order and higher orders (see, e.g.,
Refs.~\cite{Pinto:1999py,Farias:2008fs}).

Putting all terms together, we find the renormalized ETP in the OPT at
first order as given by Eq.~(\ref{VeffR}) in the text.

\section{Energy spectrum and the mass eigenvalues}
\label{appB}

By shifting the fields around their background expectation values,
$\phi_0$ and $\psi_0$ in the  Lagrangian density Eq.~(\ref{lagr2}),
the quadratic part of the Lagrangian density in the fluctuation
fields, ${\cal L}_2$, reads like
\begin{eqnarray}
{\cal L}_2 &\equiv& \frac{1}{2} \left( \phi_1, \phi_2,
\psi_1,\psi_2\right) \mathcal{M} \left(\begin{array}{cccc} \phi_{1}
  \\ \phi_{2} \\ \psi_{1} \\ \psi_{2}
\end{array}\right) 
\label{L2}
\end{eqnarray}
where $\mathcal{M}$ is the $4\times 4$ matrix of quadratic
coefficients. In the Euclidean momentum  representation it gives the
free inverse propagator matrix,
\begin{widetext}
\begin{eqnarray}
G_0^{-1}(\omega_n, {\bf p}) =  \left(\begin{array}{cccc} \omega_n^2 +
  {\bf p}^2 + M_{H,\phi}^2    & -2\mu_{\phi} \omega_n & \lambda \phi_0
  \psi_0 & 0 \\ 2\mu_{\phi} \omega_n &   \omega_n^2 + {\bf p}^2 +
  M_{G,\phi}^2    & 0 & 0 \\ \lambda \phi_0 \psi_0 & 0 &    \omega_n^2
  + {\bf p}^2 + M_{H,\psi}^2   & -2 \mu_{\psi} \omega_n\\ 0 & 0 &
  2\mu_{\psi} \omega_n &  \omega_n^2 + {\bf p}^2 + M_{G,\phi}^2 
\end{array}\right), 
\label{G0}
\end{eqnarray}
\end{widetext}
where we have defined
\begin{eqnarray}
M_{H,\phi}^2  \equiv  m^2_{\phi} -\mu_{\phi}^2 +
\frac{\lambda_{\phi}}{2}  \phi^2_0 + \frac{\lambda}{2} \psi_0^2,
\nonumber\\ M_{G,\phi}^2  \equiv  m^2_{\phi} -\mu_{\phi}^2 +
\frac{\lambda_{\phi}}{6}  \phi^2_0+ \frac{\lambda}{2} \psi_0^2, 
\label{MHMG}
\end{eqnarray}
with analogous expressions for $M_{H,\psi}$ and $M_{G,\psi}$,
replacing the fields labeling in the above expression.  The
eigenvalues of the free inverse propagator matrix $G_0^{-1}(\omega_n,
{\bf p})$, denoted here as $\omega_n^2+\varepsilon_i^2({\bf p})$,
$i=1,\ldots,4$, give the energy spectrum  (dispersion relations) for
the particles in the model. The expressions for  $\varepsilon_i^2({\bf
  p})$ are long and cumbersome, however,  the mass eigenvalues ${\cal
  M}^2_i$, when taking $\omega_n=0, |{\bf p}|=0$ in Eq.~(\ref{G0}),
are simple and they are given by
%
\begin{eqnarray}
\mathcal{M}_1^2&=& \frac{M_{H,\phi}^2+M_{H,\psi}^2}{2}  \nonumber
\\ &+&  \sqrt{\frac{\left(M_{H,\phi}^2 - M_{H,\psi}^2 \right)^2}{4}
  +\lambda^2 \phi_0^2 \psi_0^2 },
\label{MHphi}
\\  \mathcal{M}_2^2&=& M_{G,\phi}^2
\label{MGphi}
\\ \mathcal{M}_3^2&=& \frac{M_{H,\phi}^2+M_{H,\psi}^2}{2}  \nonumber
\\ &-& \sqrt{\frac{\left(M_{H,\phi}^2 - M_{H,\psi}^2 \right)^2}{4}
  +\lambda^2 \phi_0^2 \psi_0^2 },
\label{MHpsi}
\\  \mathcal{M}_4^2&=& M_{G,\psi}^2
\label{MGpsi}
\end{eqnarray}

Note that from the tree-level potential,
\begin{eqnarray}
V_{0}&=& \frac{m_{\phi}^{2}-\mu_\phi^2}{2}\phi_0^{2}+\frac{m_{\psi
  }^{2}-\mu_\psi^2}{2}\psi_0^{2} \nonumber
\\ &+&\frac{\lambda_\phi}{4!}\phi_0^{4}
+\frac{\lambda_\psi}{4!}\psi_0^{4}+\frac{\lambda}{4}
\phi_0^{2}\psi_0^{2},
\label{Vtree}
\end{eqnarray}
which when it is  minimized with respect to the fields, we obtain that
the tree-level vacuum expectation values $\tilde{\phi}_0$ and
$\tilde{\psi}_0$, are  given, respectively, by 
\begin{eqnarray}
&& \tilde{\phi}_0^2= \frac{-6\lambda_\psi (m_\phi^2-\mu_\phi^2)+18
    \lambda (m_\psi^2-\mu_\psi^2)}{\lambda_\phi \lambda_\psi-9 \lambda
    ^2},
\label{vevphi}
\\  && \tilde{\psi}_0^2= \frac{-6\lambda_\phi (m_\psi^2-\mu_\psi^2)+18
  \lambda  (m_\phi^2-\mu_\phi^2)} {\lambda_\phi\lambda_\psi-9 \lambda
  ^2}.
\label{vevpsi}
\end{eqnarray}
When substituting $\phi_0=\tilde{\phi}_0$ and $\psi_0=\tilde{\psi}_0$
in the mass eigenvalues, we can recognize that  $\mathcal{M}_1$ and
$\mathcal{M}_2$ are, respectively, the Higgs and Goldstone modes
associated  with the complex scalar field $\phi$, while
$\mathcal{M}_3$ and $\mathcal{M}_4$ are, respectively, the Higgs and
Goldstone modes associated with $\psi$.


\end{document}